\documentclass[preprint,showpacs,preprintnumbers,amsmath,amssymb]{revtex4}
\usepackage{epsfig}
\usepackage{graphicx}
\usepackage{dcolumn}
\usepackage{bm}

\def\beq{\begin{equation}}
\def\eeq{\end{equation}}
\def\eeqn{\end{equation}}
\newcommand\iden{\leavevmode\hbox{\small1\normalsize\kern-.33em1}}

\newcommand{\bea} {\begin{eqnarray}}
\newcommand{\eea} {\end{eqnarray}}

\let\jnfont=\rm
\def\NPB#1,{{\jnfont Nucl.\ Phys.\ B }{\bf #1},}
\def\PLB#1,{{\jnfont Phys.\ Lett.\ B }{\bf #1},}
\def\EPJC#1,{{\jnfont Eur.\ Phys.\ Jour.\ C }{\bf #1},}
\def\PRD#1,{{\jnfont Phys.\ Rev.\ D }{\bf #1},}
\def\PRL#1,{{\jnfont Phys.\ Rev.\ Lett.\ }{\bf #1},}
\def\MPLA#1,{{\jnfont Mod.\ Phys.\ Lett.\ A }{\bf #1},}
\def\JPG#1,{{\jnfont J.\ Phys.\ G }{\bf #1},}
\def\CTP#1,{{\jnfont Commun.\ Theor.\ Phys.\ }{\bf #1},}
\def\JHEP#1,{{\jnfont JHEP \ }{\bf #1},}
\def\NPPS#1,{{\jnfont Nucl.\ Phys.\ Proc.\ Suppl.\ }{\bf #1},}
\def\CPC#1,{{\jnfont Computl.\ Phys.\ Commun.\ }{\bf #1},}
\def\CPL#1,{{\jnfont Chin.\ Phys.\ Lett. }{\bf #1},}
\def\AJS#1,{{\jnfont Astrophys.\ J.\ Suppl. }{\bf #1},}
\def\PR#1,{{\jnfont Phys.\ Rept. }{\bf #1},}
\def\AP#1,{{\jnfont Astropart.\ Phys. }{\bf #1},}
\def\EPL#1,{{\jnfont Europhys.\ Lett. }{\bf #1},}
\def\FP#1,{{\jnfont Fortsch.\ Phys. }{\bf #1},}

\begin{document}

\title{\ \\[10mm] Top quark forward-backward asymmetry and charge asymmetry in
                  left-right twin Higgs model
                  }

\author{Lei Wang$^1$, Lei Wu$^{2}$, Jin Min Yang$^{2}$}

\affiliation{
$^1$ Department of Physics, Yantai University, Yantai 264005, China\\
$^2$ State Key Laboratory of Theoretical Physics, \\
     Institute of Theoretical Physics, Academia Sinica,
             Beijing 100190, China}


\begin{abstract}
In order to explain the Tevatron anomaly of the top quark
forward-backward asymmetry $A_{FB}^t$ in the left-right twin Higgs
model, we choose to give up the lightest neutral particle of
$\hat{h}$ field as a stable dark matter candidate. Then a new Yukawa
interaction for $\hat{h}$ is allowed, which can be free from the
constraint of same-sign top pair production and contribute sizably
to $A_{FB}^t$. Considering the constraints from the production rates
of the top pair ($t\bar t$), the top decay rates and $t\bar{t}$
invariant mass distribution, we find that this model with such new
Yukawa interaction can explain $A_{FB}^t$ measured at the Tevatron
while satisfying the charge asymmetry $A_{C}^t$ measured at the LHC.
Moreover, this model predicts a strongly correlation between
$A_{C}^t$ at the LHC and $A_{FB}^t$ at the Tevatron, i.e., $A_{C}^t$
increases as $A_{FB}^t$ increases.
\end{abstract}


\maketitle

\section{Introduction}
The forward-backward asymmetry $A_{FB}^t$ in top quark pair
production has been measured by the two experimental groups at the
Tevatron. The CDF measured the asymmetry in the $\ell+j$ channel and
obtained $A_{FB}^t(CDF)= 0.158\pm0.074$ \cite{11100014-1}, which is
nearly consistent with the D0 result $A_{FB}^t(D0) = 0.19\pm0.065$
\cite{11100014-2}. These results exceed SM prediction, $A_{FB}^t(SM)
= 0.058\pm 0.009$, which arises from NLO QCD diagrams
\cite{11084005-4¨C7}. Including the resummation of soft-gluon
emission at NNLL, Ref. \cite{11066051} gives the currently most
precise QCD prediction, $0.072^{+0.011}_{-0.007}$.  The CDF also
reported an abnormally large value of $A_{FB}^t$  for $m_{t\bar{t}}>
450$ GeV \cite{11100014-1}, which, however, is not confirmed by D0
collaboration \cite{11100014-2}.

To explain $A_{FB}^t$, various attempts have been tried, such as via
the $s$-channel exchange of an axi-gluon \cite{afb-s-channel} or the
$t$-channel exchange of $Z'$, $W'$ and a scalar
\cite{afb-tu-channel,11040083,11040083-19,11080998,afbdefi,11074350,09113237}
or through an effective model-independent way
\cite{afb-eft,afb-eft-2}. In this work we will try to explain
$A_{FB}^t$ in the framework of the left-right twin Higgs model
(LRTH) \cite{twinhiggs,lrth,phlrth}. In this model, a discrete
left-right symmetry ensures the absence of one-loop quadratic
divergence of the SM Higgs mass, which emerges as a pseudo-Goldstone
boson once a global symmetry is spontaneously broken. The resulting
Higgs boson mass is naturally around the electroweak scale when the
cut-off scale of the theory is around 5-10 TeV. In the original
LRTH,  the lightest neutral particle of $\hat{h}$ field is stable
and thus can be a candidate for weakly interacting massive particle
(WIMP) dark matter \cite{lrthdm}. We found that the original LRTH
does not contribute to $A_{FB}^t$ sizably, so we choose to give up
the dark matter candidate. Then a new Yukawa interaction for
$\hat{h}$ is allowed, which is found to contribute sizably to
$A_{FB}^t$.

In our analysis we will consider the following observables:
\begin{itemize}
\item[(1)] $A_{FB}^t$ in the $t\bar{t}$ rest
frame at Tevatron, which is defined by \cite{afbdefi}
\beq
A_{FB}^t=A_{FB}^{NP}\times R + A_{FB}^{SM}\times (1-R) \label{afbgongsi}
\eeq
where $A_{FB}^{SM}=0.058$ is the asymmetry in the SM, and
\begin{eqnarray}
&& A_{FB}^{NP}=\frac{\sigma^{NP}(\Delta y >0)
-\sigma^{NP}(\Delta y <0)}{\sigma^{NP}(\Delta y >0)+\sigma^{NP}(\Delta y <0)},\\
&& R=\frac{\sigma^{NP}}{\sigma^{SM}+\sigma^{NP}}
\label{afbgongsi3}\end{eqnarray} are the asymmetry induced by the
new physics and the fraction of the new physics contribution to the
total cross section, respectively. $\Delta y$ is the rapidity
difference between a top and an anti-top.

\item[(2)] The charge asymmetry of $t\bar{t}$ production at LHC, defined by
\beq
A_{C}^t=A_{C}^{NP}\times R + A_{C}^{SM}\times (1-R)
\eeq
where $A_{FB}^{SM}=0.013$ is  the asymmetry in the SM \cite{lacafbdefi},
and
\begin{eqnarray}
&& A_{C}^{NP}\equiv\frac{\sigma^{NP}(|\eta_t|>|\eta_{\bar{t}}|)
-\sigma^{NP}(|\eta_t|<|\eta_{\bar{t}}|)}
{\sigma^{NP}(|\eta_t|>|\eta_{\bar{t}}|)+\sigma^{NP}(|\eta_t|<|\eta_{\bar{t}}|)},\\
&&  R=\frac{\sigma^{NP}}{\sigma^{SM}+\sigma^{NP}}
\end{eqnarray}
are the asymmetry induced by the new physics and the fraction of the
new physics contribution to the total cross section, respectively.
$\eta_t$ and $\eta_{\bar{t}}$ are respectively the pseudo-rapidity
of top and anti-top quark in the laboratory frame. This asymmetry
reflects that the top quarks on average are more boosted than the
anti-top quarks, which is sensitive to new physics beyond the SM
\cite{afb-eft-2,ac-np}. The CMS collaboration has recently measured
the quantity with an integrated luminosity of 1.09 $fb^{-1}$ and
obtained $A_C^t=-0.016\pm0.030^{+0.010}_{-0.019}$, which is
consistent with the SM prediction \cite{lacafbdefi}. The
uncertainties of the ATLAS measurement of the charge asymmetry are
of similar size with respect to the CMS result \cite{atlas-ac}.

\item[(3)] The $t\bar{t}$ total production cross sections at
Tevatron and LHC. The current cross section measured at Tevatron is
$\sigma^{exp}=7.50\pm0.48$ pb for $m_t=172.5$ GeV
\cite{11070841-46}, while the SM cross section is
$\sigma^{SM}=7.46^{+0.66}_{-0.80}$ pb from \cite{11070841-47} and
$\sigma^{SM}=6.30\pm0.19^{+0.31}_{-0.23}$ pb from
\cite{11070841-48}. The $t\bar{t}$ total production cross section
measured recently at LHC with the center of mass energy 7 TeV is
$\sigma^{exp}=176\pm5^{+13}_{-10}\pm7$ pb from ATLAS \cite{11101027}
and $\sigma^{exp}=168\pm18\pm14\pm7$ pb from CMS \cite{11055661},
while the SM cross section is
$\sigma^{SM}=165.80^{+4.44}_{-6.99}\pm9.10\pm11.6$ pb from
\cite{11070841-47} and
$\sigma^{SM}=157.92^{+7.79}_{-8.88}\pm8.67\pm11.9$ pb from
\cite{top11008pas-9}. Here, we conservatively require
$-0.12<\frac{\sigma^{NP}}{\sigma^{SM}}<0.3$ for the Tevatron and
$-0.25<\frac{\sigma^{NP}}{\sigma^{SM}}<0.25$ for the LHC.

\item[(4)] The top quark can decay into a light quark and a scalar particle
for the scalar mass is light enough. The measurement of the total
top width is $\Gamma_t^{exp}=1.99^{+0.69}_{-0.55}$ GeV
\cite{11074350-49}, and is in agreement with the SM value
$\Gamma_t^{SM}=1.3$ GeV, which sets a limit on the partial width of
any new decay mode.
\end{itemize}

Finally, we will discuss the constraints from the experimental data
of $t\bar{t}$ invariant mass distribution and single top quark
production.

This work is organized as follows. In Sec. II, we briefly review the
left-right twin Higgs model and then introduce a new Yukawa
interaction for $\hat{h}$. In Sec. III, we study the top quark
observables mentioned above, and focus on the top quark
forward-backward asymmetry at Tevatron and charge asymmetry at LHC
under the constraints of the other observables. Finally, we give our
conclusion in Sec. IV.

\section{LRTH model with new Yukawa interaction}

The LRTH model \cite{lrth,phlrth} has a global symmetry $U(4)\times
U(4)$ with a gauged $SU(2)_L\times SU(2)_R\times U(1)_{B-L}$
subgroup. The twin symmetry is identified as a left-right symmetry
with interchanging L and R, which implies that the gauge couplings
of $SU(2)_L$ and $SU(2)_R$ are identical ($g_{2L}=g_{2R}=g_2$).

A pair of Higgs fields, $H$ and $\hat{H}$, are introduced, which
transform as $(\textbf{4},\textbf{1})$ and
$(\textbf{1},\textbf{4})$ respectively under the global symmetry.
They can be written as
\begin{equation}
H=\left(
\begin{tabular}{c}
$H_L$\\
$H_R$
\end{tabular}
\right),\ \ \ \ \ \hat{H}=\left(
\begin{tabular}{c}
$\hat{H}_L$\\
$\hat{H}_R$
\end{tabular}
\right),
\end{equation}
where $H_{L,R}$ and $\hat{H}_{L,R}$ are two component objects which
are charged under $SU(2)_L\times SU(2)_R\times U(1)_{B-L}$ as
\begin{equation}
    H_L {\rm \ and\ }\hat{H}_L: ({\bf 2},{\bf 1},1); \ \ \
    H_R {\rm \ and\ }\hat{H}_R: ({\bf 1},{\bf 2},1).
\end{equation}
The SM-like Higgs doublet $h=(h^+, h^0)^T$ and the new doublet
 $\hat{h}=(\hat{h}^+, \hat{h}^0)^T$ reside in $H_L$ and
 $\hat{H}_L$, respectively.

 Each Higgs acquires a non-zero VEV as
\begin{equation}\label{eq:vev1}
    <H> ~=~ (~0~~0~~0~~f~)^T,\;\;\;\;\; <\hat{H}> ~=~ (~0~~0~~0~~
  \hat{f}~)^T,
\end{equation}
which breaks one of the $U(4)$ to $U(3)$ and yields seven
Nambu-Goldstone bosons. The gauge symmetry $SU(2)_L\times
SU(2)_R\times U(1)_{B-L}$ is broken down to $U(1)_{EM}$, and six out
of the fourteen Goldstone bosons are respectively eaten by the SM
gauge bosons $W$ and $Z$, and additional gauge boson $W_H$ and $Z_H$
with masses of a few TeV. In addition to the SM-like Higgs, we are
left with the two neutral pseudoscalar $\phi^0$ and $\hat{A}$, one
neutral scalar $\hat{S}$, and the charged scalar $\phi^\pm$ and
$\hat{h}^\pm$. Here $\hat{S}$ and $\hat{A}$ are from
$\hat{h}^0=(\hat{S}+i\hat{A})/\sqrt{2}$.

The SM quarks and leptons are charged under $SU(2)_L\times SU(2)_R\times
U(1)_{B-L}$ as
\begin{eqnarray}
    L_{L\alpha}=-i\left(\begin{array}{c}~\nu_{L\alpha}
\\l_{L\alpha}\end{array}\right):
({\bf 2},{\bf 1},-1), \ \ \ \ \ &&
    L_{R\alpha}=\left(\begin{array}{c}~\nu_{R\alpha}
\\l_{R\alpha}\end{array}\right):
({\bf 1},{\bf 2},-1),\nonumber\\
    Q_{L\alpha}=-i\left(\begin{array}{c} u_{L\alpha}
\\ d_{L\alpha}\end{array}\right):
({\bf 2},{\bf 1},1/3),\ \ \ \ \ &&
    Q_{R\alpha}=\left(\begin{array}{c}u_{R\alpha}
\\d_{R\alpha}\end{array}\right):({\bf 1},{\bf 2},1/3)
\end{eqnarray} with $\alpha$
being the family index.

After the doublet $h$ residing in $H_L$ acquires the VEV,
$v\approx246$ GeV, the masses of the first two generation quarks and
bottom quark can be obtained from \cite{phlrth}
\begin{equation}
    {\cal L}_{Y}=\frac{\lambda_u^{\alpha\beta}}
{\Lambda}(\bar{Q}_{L\alpha}\tau_2 H_L^*)(H_R^T\tau_2{Q}_{R\beta})
+\frac{\lambda_d^{\alpha\beta}}{\Lambda}(\bar{Q}_{L\alpha}
H_L)(H_R^{\dagger}{Q}_{R\beta}) + h.c.,
\label{Yukawa2}
\end{equation}
where $\tau_2=i\sigma_2$ ($\sigma_2$ is Pauli matrix). The Yukawa
interaction of leptons is similar to Eq. (\ref{Yukawa2}).

In order to explain the top quark forward-backward asymmetry at
Tevatron, we add the new Yukawa interaction:
\begin{equation}
    {\cal L}_{q}=\frac{y_u^{\alpha\beta}}
{\Lambda}(\bar{Q}_{L\alpha}\tau_2
\hat{H}_L^*)(\hat{H}_R^T\tau_2{Q}_{R\beta})
+\frac{y_d^{\alpha\beta}}{\Lambda}(\bar{Q}_{L\alpha}
\hat{H}_L)(\hat{H}_R^{\dagger}{Q}_{R\beta}) + h.c..
\label{Yukawahat}
\end{equation}
Since the VEV of $\hat{H}_L$ equals to zero, the interaction can not
produce the mass term of SM quark. With the mass eigenstates and the
expressions of $\hat{H}_L$ and $\hat{H}_R$ shown in \cite{phlrth},
we then obtain the following couplings \bea {\cal
L}_{q}=&-&\frac{\hat{f}}
{\Lambda}\left(\hat{h}^{0*}~(X_u)_{\alpha\beta}~ \bar{u}_L^{\alpha}
u_R^{\beta}-\hat{h}^- ~(V_{CKM}^\dag
X_u)_{\alpha\beta}~\bar{d}_L^{\alpha} u_R^{\beta}
\right) \nonumber\\
&-&\frac{\hat{f}} {\Lambda}\left(\hat{h}^0 ~(X_d)_{\alpha\beta} ~\bar{d}_L^{\alpha} d_R^{\beta}
+\hat{h}^+~(V_{CKM}X_d)_{\alpha\beta}~\bar{u}_L^{\alpha} d_R^{\beta}
\right) +h.c..
 \label{yukhat}
\eea

To satisfy the constraints from the flavor processes and electroweak data, we take two cases
for the mixing matrixes $X_u$ and $X_d$
(the detailed analysis was given in \cite{11074350}):
\begin{itemize}
\item[(i)] Case I: $(X_u)_{\alpha1}=\kappa_1(V_{CKM})_{\alpha3}$, $(X_u)_{\alpha2}=0$,
$(X_u)_{\alpha3}=0$ and $(X_d)_{\alpha\beta}=0$. From Eq.
(\ref{yukhat}), we can obtain the coupling \bea {\cal
L}_{q}&=&-\frac{\kappa_1\hat{f}}
{\Lambda}\left((V_{CKM})_{\alpha3}~\hat{h}^{0*}~ \bar{u}_L^{\alpha}
u_R-\hat{h}^- ~\bar{b}_L u_R
\right)+h.c.\nonumber \\
&=&-2y_1\left((V_{CKM})_{\alpha3}~\hat{h}^{0*}~ \bar{u}_L^{\alpha}
u_R-\hat{h}^- ~\bar{b}_L u_R \right)+h.c.
 \label{yukhat1}
\eea with $y_1=\frac{\kappa_1\hat{f}} {2\Lambda}$.

\item[(ii)]  Case II: $(X_u)_{\alpha\beta}=0$ and $(X_d)_{\alpha\beta}=0$ except for
 $(X_d)_{31}=\kappa_2$.
From Eq. (\ref{yukhat}), we can obtain the coupling \bea {\cal
L}_{q}&=&-\frac{\kappa_2\hat{f}} {\Lambda}\left(\hat{h}^{0}~
\bar{b}_L d_R+(V_{CKM})_{\alpha3}~\hat{h}^+ ~\bar{u}_L^{\alpha} d_R
\right)+h.c.\nonumber \\
&=&-2y_2\left(\hat{h}^{0}~ \bar{b}_L
d_R+(V_{CKM})_{\alpha3}~\hat{h}^+ ~\bar{u}_L^{\alpha} d_R
\right)+h.c.
 \label{yukhat2}
\eea with $y_2=\frac{\kappa_2\hat{f}} {2\Lambda}$.
\end{itemize}

The cut-off scale $\Lambda$ is typically taken to be $4 \pi f$ with
$f$ being as low as 500 GeV. Sometime $\Lambda=2\pi f$ is also
considered \cite{phlrth}. The scale $\hat{f}$ can be determined from
the electroweak symmetry breaking condition. At a rough estimate,
$\hat{f}$ is five times as $f$ or more \cite{phlrth,0701071}. For
Case I (Case II), $\hat{S}$ and $\hat{A}$ from
$\hat{h}^0=\frac{\hat{S}+i\hat{A}}{\sqrt{2}}$ ($\hat{h}^\pm$) can
contribute to the top quark forward-backward asymmetry at the
Tevatron via the $t$-channel exchange of such a scalar. This also
implies that $\hat{S}$ or $\hat{A}$ can no longer be the candidate
for the WIMP dark matter.

The Coleman-Weinberg potential and the soft left-right symmetry
breaking terms (the so-called $\mu$-term) can give masses for
$\hat{h}^\pm$ and $\hat{h}^0$ as \cite{phlrth}
\begin{eqnarray}
&& m^2_{\hat S}=m^2_{\hat A}=m^2_{\hat h^0}=\frac{3}{16\pi^2}\Big[\frac{g_2^2}{2} ({\mathcal
Z}(m_W)-{\mathcal Z}(m_{W_H})) \nonumber\\
&&~~~~~~~~~~~~~~ +\frac{2g_1^2+g_2^2}{4}\frac{m^2_{W_H}-m^2_W}{m^2_{Z_H}-m^2_Z}
({\mathcal Z}(m_Z)-{\mathcal Z}(m_{Z_H}))\Big]
+\mu^2_r\frac{f}{\hat f}\cos x+\hat \mu^2, \\
&& m^2_{\hat h^\pm} \simeq m^2_{\hat h^0}, \label{h2h1mass}
\end{eqnarray}
where ${\mathcal Z}(x)=-x^2(\ln\frac{\Lambda^2}{x^2}+1)$. The last
two terms are from the $\mu$-term. We neglect the small mass
splitting between $\hat{h}^0$ and $\hat{h}^\pm$ due to the
electromagnetic interactions. Note that $\hat{\mu}^2$ could have
either sign, which can allow us to vary the masses of $\hat{h}^0$
and $\hat{h}^\pm$ as a free parameter.

Note that, due to an additional phase factor $i$ in the Yukawa
coupling of $\hat{A}$, the contributions of $\hat{S}$ and $\hat{A}$
to the same-sign top pair productions are destructive and such
contributions can be even canceled for the degeneracy masses of
$\hat{S}$ and $\hat{A}$. Thus, the LRTH with such new Yukawa
interaction can be free from the strong constraints from $tt$
production rate reported by CMS collaboration, $\sigma(tt)<17$ pb at
95\% C. L. \cite{11062142}.

In fact, we still can introduce a parity in the model under which
$\hat{H}$ is odd while all the other fields are even. The
non-renormalizable interaction of Eq. (\ref{Yukawahat}) is invariant
under this parity. This parity can forbid the renormalizable
interaction between $\hat{H}$ and fermions, especially the top
quark. The top quark mass can still be obtained from the
renormalizable interaction shown in the original LRTH \cite{phlrth}.

\section{calculations and discussions}
In our calculations, we take $m_t=172.5$ GeV and use the parton
distribution function CTEQ6L \cite{cteq} with renormalization scale
and factorization scale $\mu_R = \mu_F=m_t$. We assume that the
K-factors are universal, so that the QCD correction effects are
canceled in the ratios of $\sigma^{NP}/\sigma^{SM}$ and
$\sigma^{NP}/(\sigma^{SM}+\sigma^{NP})$, and they are the same at LO
and NLO.

\begin{figure}[tb]
 \epsfig{file=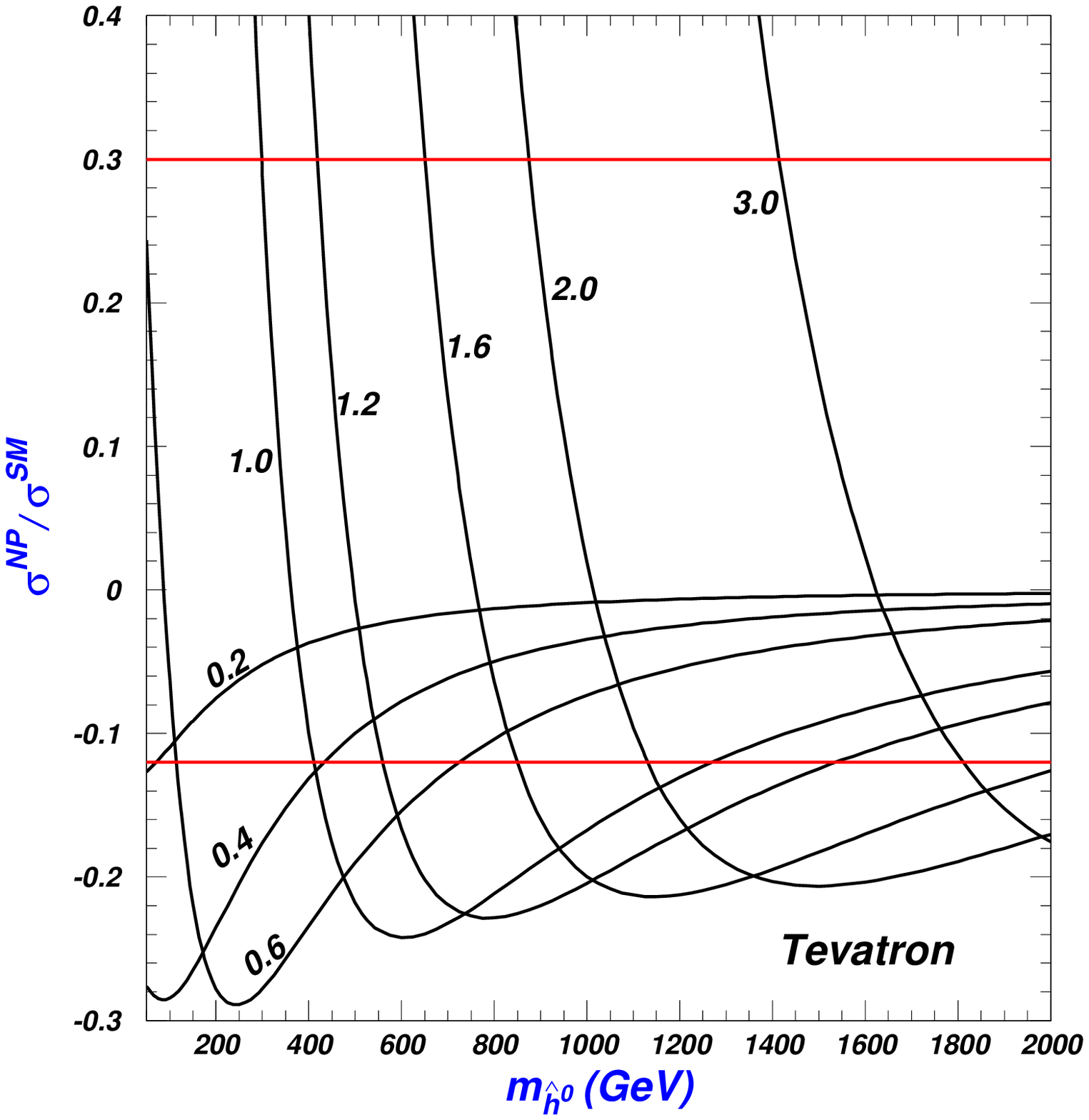,height=5.6cm}
 \epsfig{file=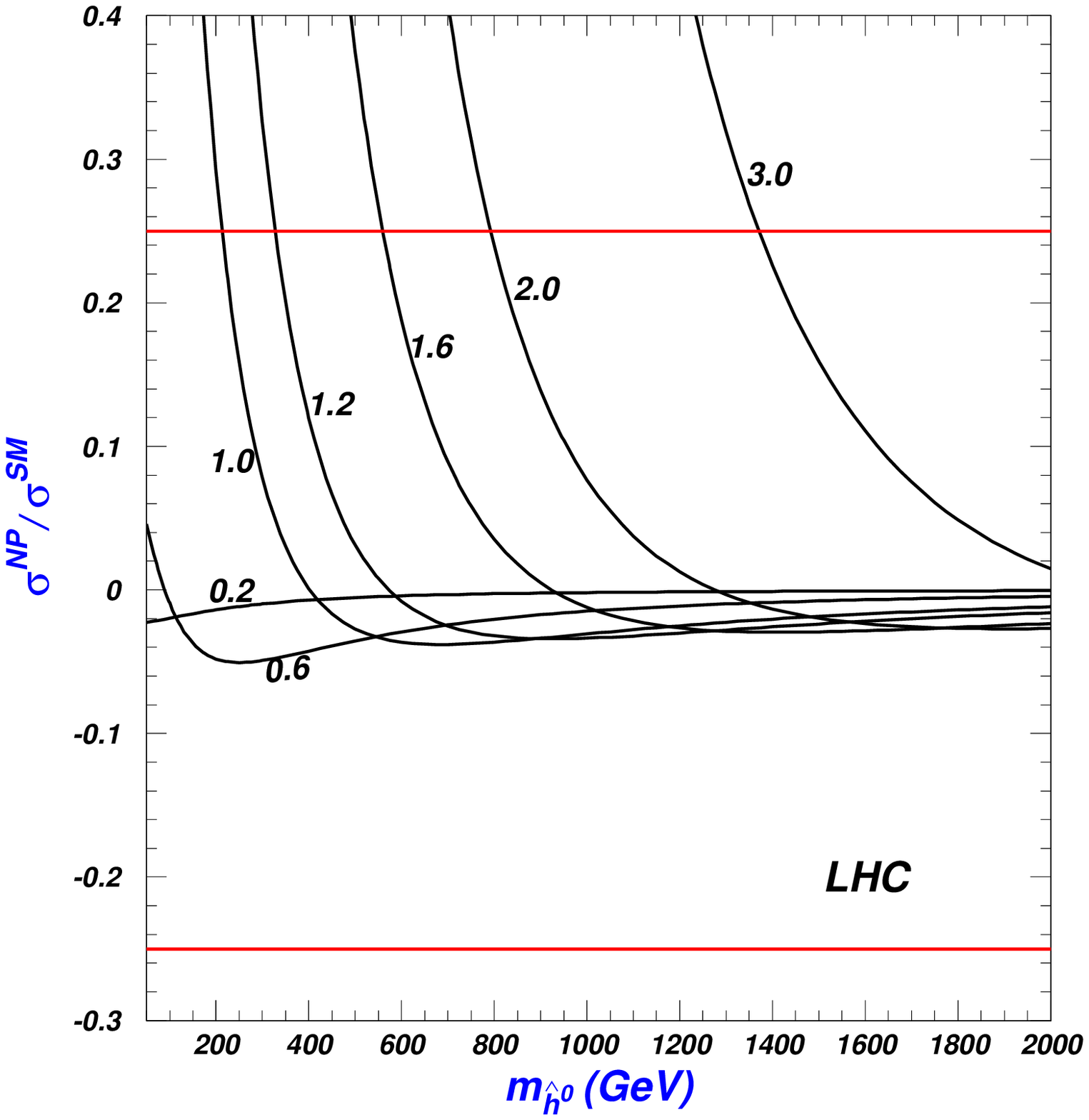,height=5.6cm}
 \epsfig{file=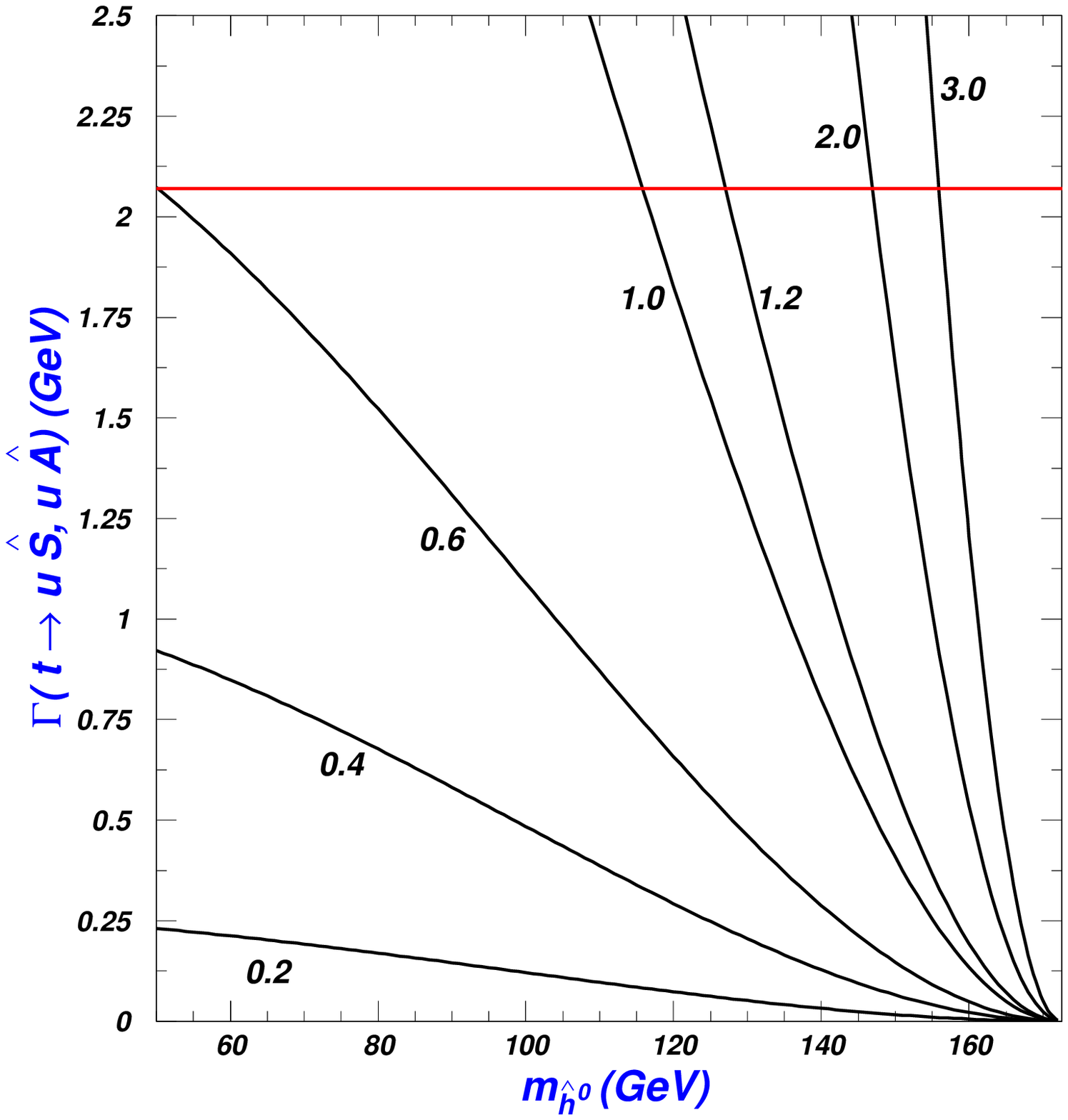,height=5.6cm}
\vspace{-1.2cm} \caption{For Case I, the new physics contributions
to $t\bar{t}$ production rates (normalized to SM values) and the
decay width of $t\to u\hat{S},~u\hat{A}$ versus $m_{\hat{h}^0}$. The
numbers on the curves denote the Yukawa coupling $y_1$. The
horizontal lines show the $2\sigma$ limits from the corresponding
experimental data.} \label{caseI}
\end{figure}

\subsection{Case I: $\hat{S}$ and $\hat{A}$}
For Case I, the matrix elements $M$ of the process $u(p_1)\bar{u}(p_2)\to t(k_1)\bar{t}(k_2)$,
including
the SM, new scalar $\hat{S}$ and $\hat{A}$ contributions, can be written as ref. \cite{09113237}
\beq
\sum|M|^2=\frac{16g_s^4}{s^2}(t_t^2+u_t^2+2sm_t^2)+32g_s^2y^2\frac{sm_t^2+t_t^2}{st_{\hat{h}^0}}
+36\frac{y^4t_t^2}{t^2_{\hat{h}^0}},
\label{amph0}\eeq
where $s=(p_1+p_2)^2$, $t=(p_1-k_1)^2$, $u=(p_1-k_2)^2$, $t_t=t-m_t^2$,
 $t_{\hat{h}^0}=t-m^2_{\hat{h}^0}$, $y=\sqrt{2}y_1$.

\begin{figure}[tb]
 \epsfig{file=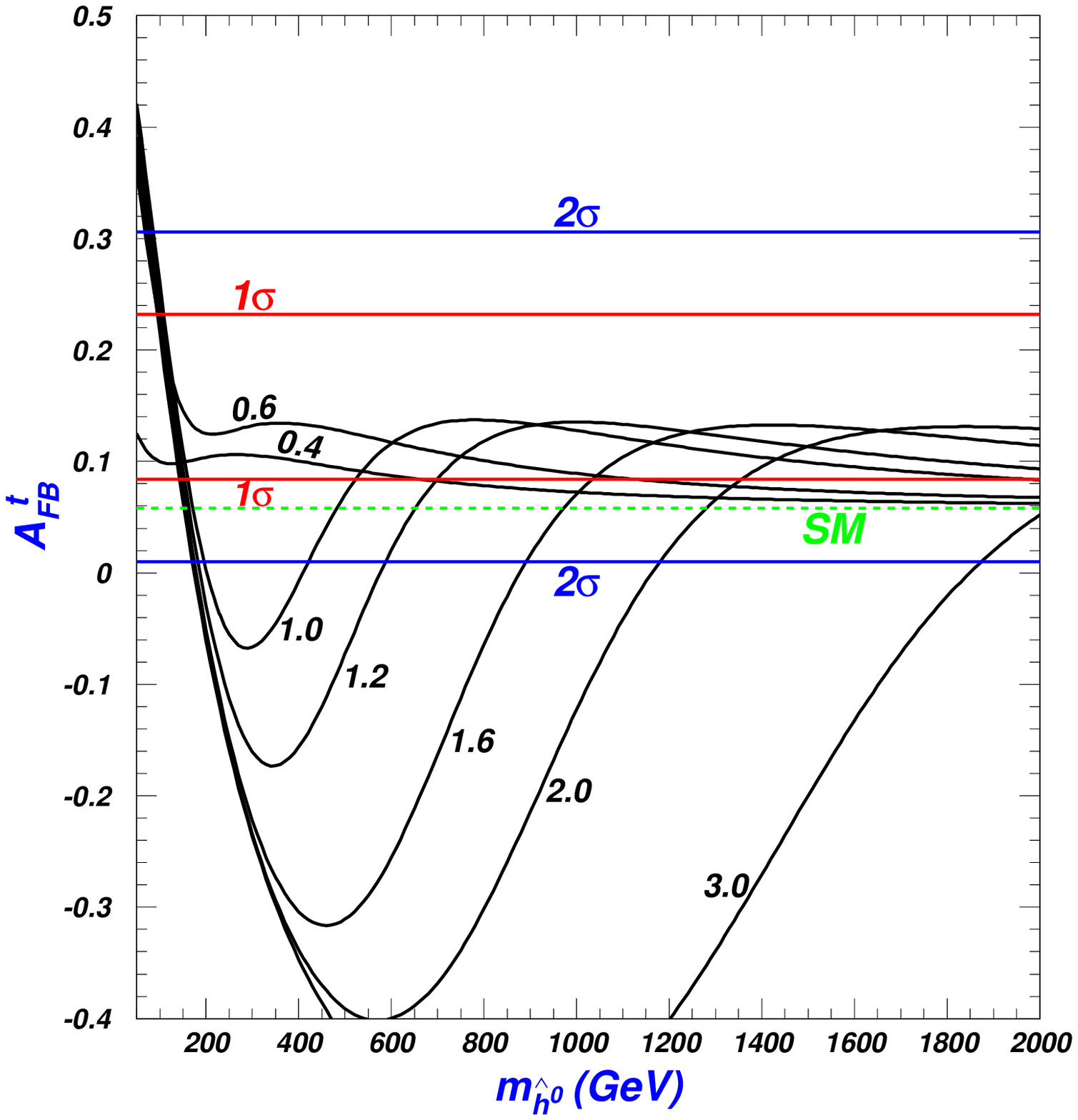,height=7.5cm}
 \epsfig{file=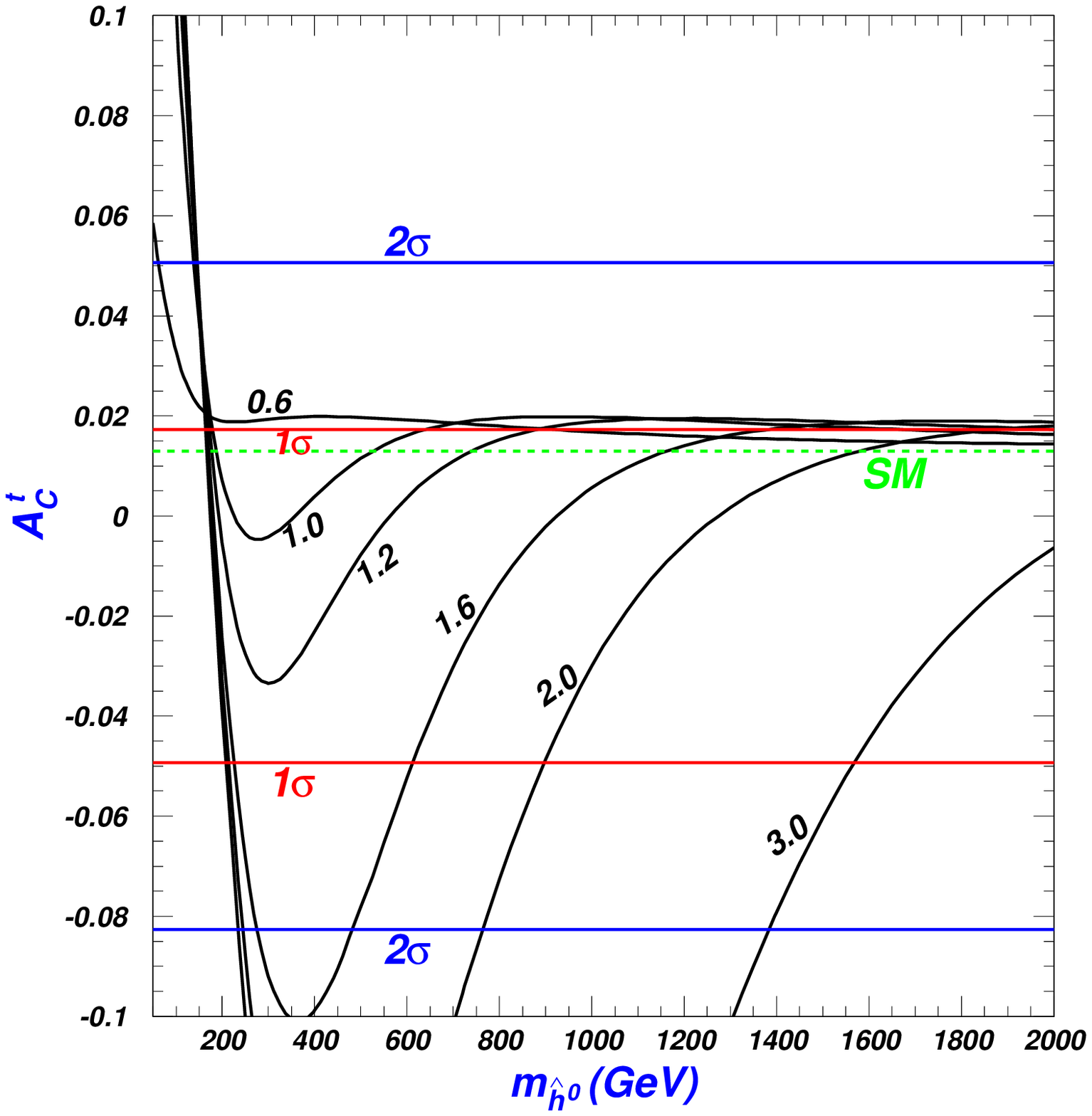,height=7.5cm}
\vspace{-.5cm} \caption{For Case I, the top forward-backward
asymmetry $A_{FB}^t$ at Tevatron and charge asymmetry $A_{C}^t$ at
LHC versus $m_{\hat{h}^0}$. The dash lines denote the SM
predictions. The horizontal lines show the $1\sigma$ and $2\sigma$
ranges from the corresponding experimental data.} \label{Iafbc}
\end{figure}

\begin{figure}[tb]
 \epsfig{file=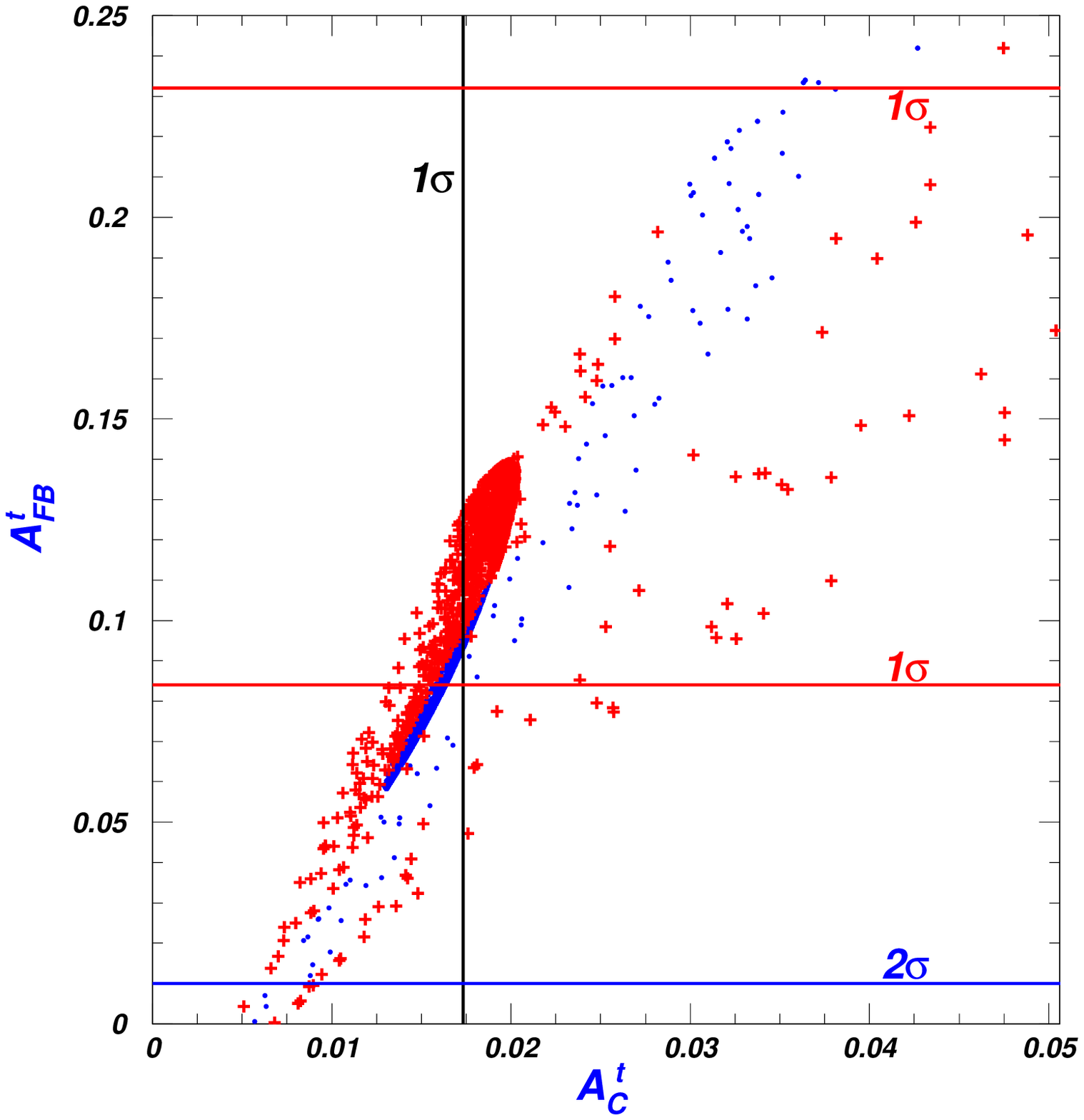,height=6.0cm}
\vspace{2.0cm} \caption{For Case I, the top quark forward-backward
asymmetry $A_{FB}^t$ at Tevatron versus charge asymmetry
 $A_{C}^t$ at LHC. The bullets (blue) and crosses (red) are
respectively allowed and excluded by the three observables shown in
Fig. \ref{caseI}. The horizontal lines show the $1\sigma$ and
$2\sigma$ lower limits from the experimental data of $A_{FB}^t$ at
Tevatron. The vertical line shows the $1\sigma$ upper limit from the
experimental data of $A_{C}^t$ at LHC.} \label{afblhc}
\end{figure}

In Fig. \ref{caseI}, we plot respectively the new physics
contributions to $t\bar{t}$ production at Tevatron and LHC
normalized to SM one, and the decay width of $t\to
u\hat{S},~u\hat{A}$ for Case I. We can find that the contributions
of $\hat{S}$ and $\hat{A}$ to the $t\bar{t}$ cross section can be
positive or negative, which depends on the coupling constant $y_1$
and their masses. Since the process $gg\to t\bar{t}$ dominates the
$t\bar{t}$ cross section at LHC and the contributions of $\hat{S}$
and $\hat{A}$ are from the process $u\bar{u}\to t\bar{t}$, the
magnitude of $\frac{\sigma^{NP}}{\sigma^{SM}}$ at LHC is smaller
than that of Tevatron. The $t\bar{t}$ cross section measured at
Tevatron gives the most constraint on the parameters $y_1$ and
$m_{\hat{h}^0}$. For example, the measurement value requires
$m_{\hat{h}^0}$ to be larger than 1200 GeV (2000 GeV) in addition to
the narrow intermediate region for $y_1 =1.0$ (1.6). The $t\bar{t}$
cross section measured at LHC and top quark decay can hardly give
further constraints.

In Fig. \ref{Iafbc}, we plot the top quark forward-backward asymmetry $A_{FB}^t$ at Tevatron and charge asymmetry
 $A_{C}^t$ at LHC for Case I. We can see that $A_{FB}^t$ can be enhanced sizably for the very low values of
 $m_{\hat{h}^0}$, be over 0.1 for the large ones and be negative in the intermediate region.
For the large region of $m_{\hat{h}^0}$, the left panel of Fig.
\ref{caseI} shows that $\frac{\sigma^{NP}}{\sigma^{SM}}$ is
negative, which can play a positive role in enhancing the $A_{FB}^t$
according to its definition shown in Eq. (\ref{afbgongsi}) and Eq.
(\ref{afbgongsi3}). The dependence of $A_{C}^t$ on $y_1$ and
$m_{\hat{h}^0}$ is similar to that of $A_{FB}^t$, which is within
$1\sigma$ range in the large parameter spaces.

In Fig. \ref{afblhc}, we scan the following parameter space,
$$0.1 \leq y_1 \leq 1.0,~~~100~ GeV \leq m_{\hat{h}^0} \leq 2000~GeV,$$ and
plot $A_{FB}^t$ versus $A_{C}^t$ under the constraints of the three
observables shown in Fig. \ref{caseI}. We find that $A_{FB}^t$ and
$A_{C}^t$ have direct correlation, and the former always increases
as increasing of the latter. The $A_{FB}^t$ can be explained to
within $1\sigma$ and reach 0.1 for $A_{C}^t$ remains within
$1\sigma$.  For $A_{C}^t$ is in the range of $1\sigma$ and
$2\sigma$, $A_{FB}^t$ can reach 0.24. If the future more precision
measurement at LHC shows that $A_{C}^t$ is smaller than 0.0125, the
model will lose its spirit of producing a large $A_{FB}^t$ at the
Tevatron.

\subsection{Case II: $\hat{h}^\pm$}
For Case II, the matrix elements $M$ of the process $d(p_1)\bar{d}(p_2)\to t(k_1)\bar{t}(k_2)$,
 including the SM and  $\hat{h}^+$ contributions, is the same
as Eq. (\ref{amph0}), but replacing $m_{\hat{h}^0}$ and $y_1$ with $m_{\hat{h}^+}$ and $y_2$.

\begin{figure}[tb]
 \epsfig{file=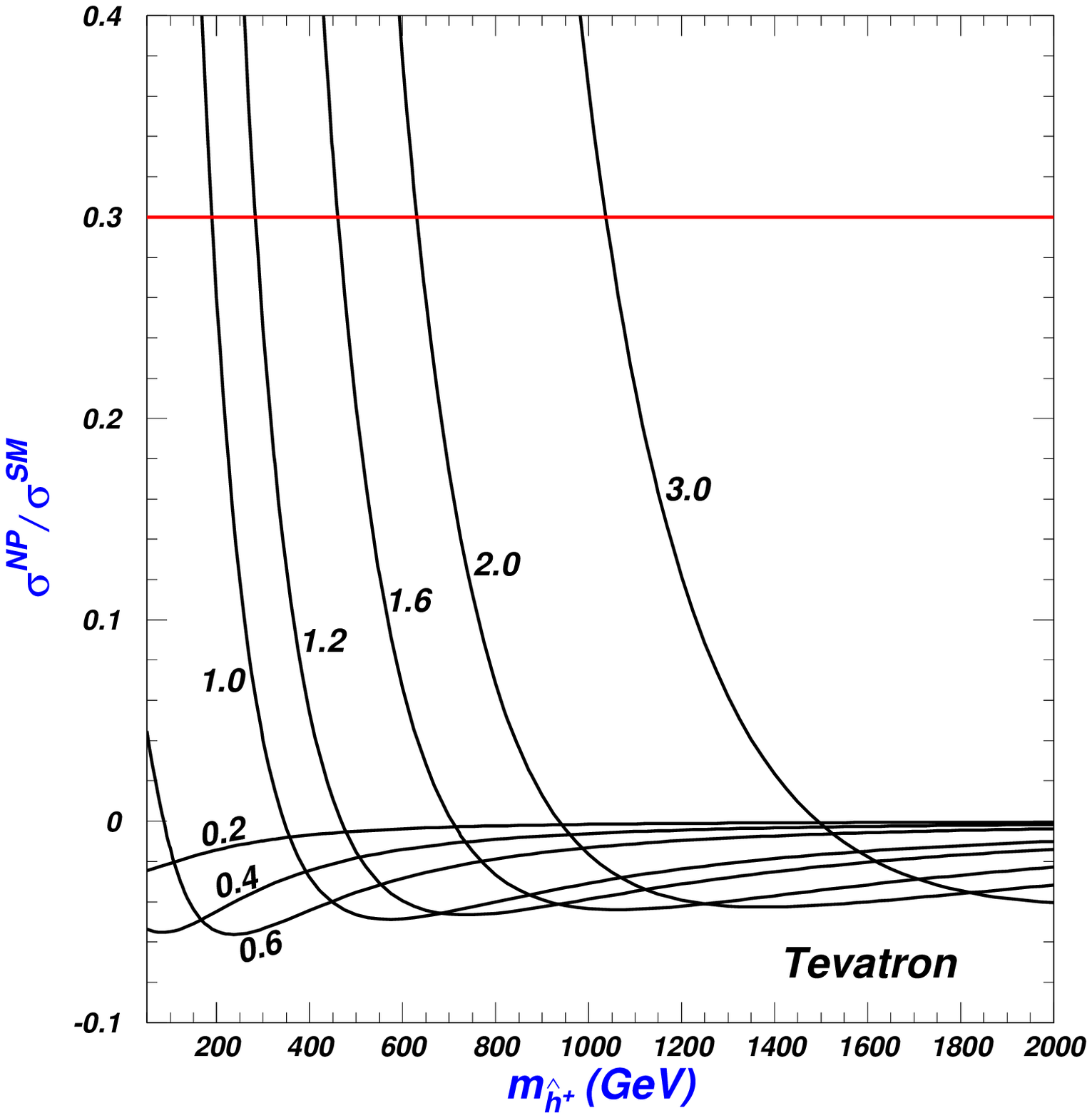,height=5.6cm}
 \epsfig{file=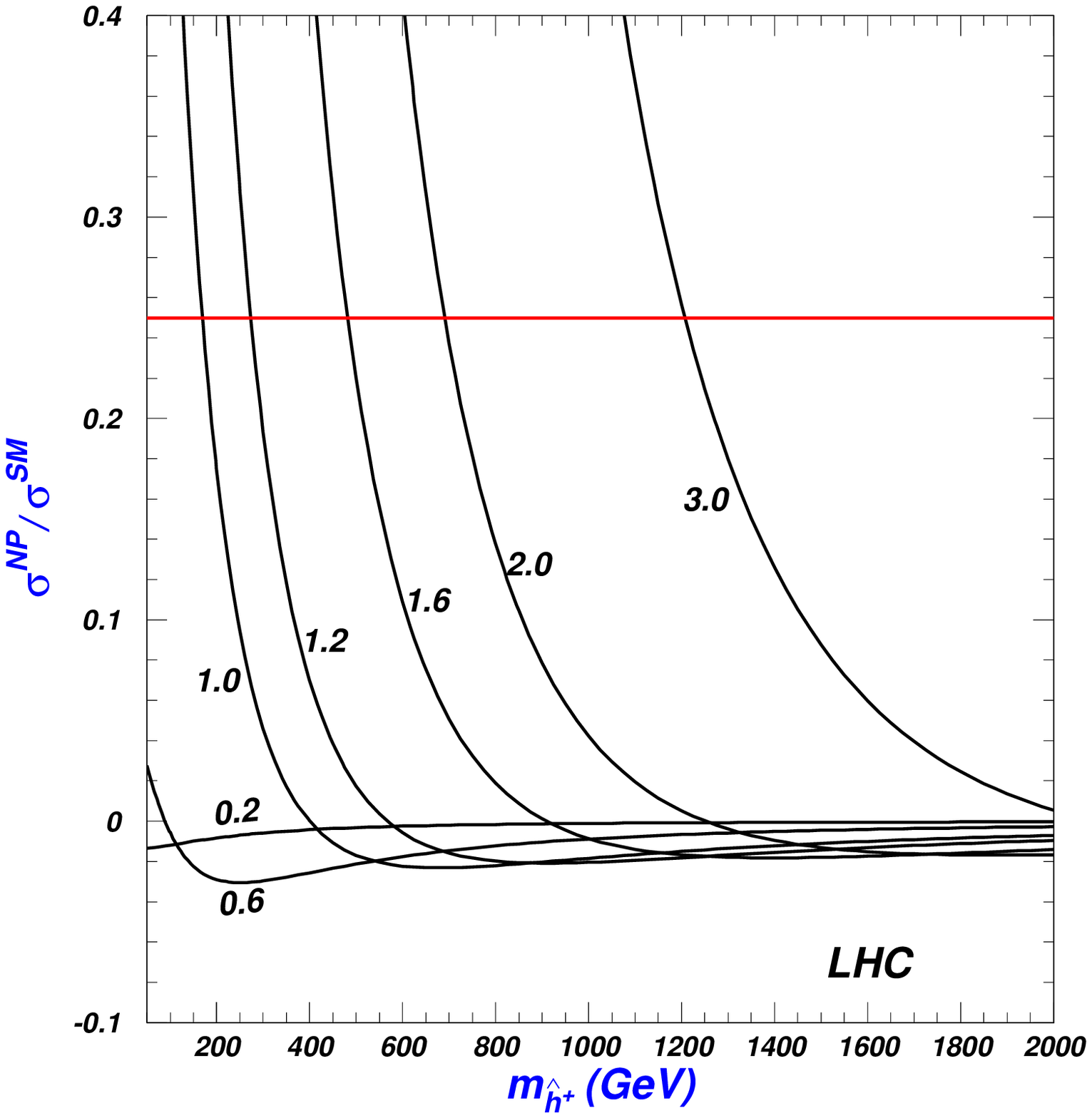,height=5.6cm}
 \epsfig{file=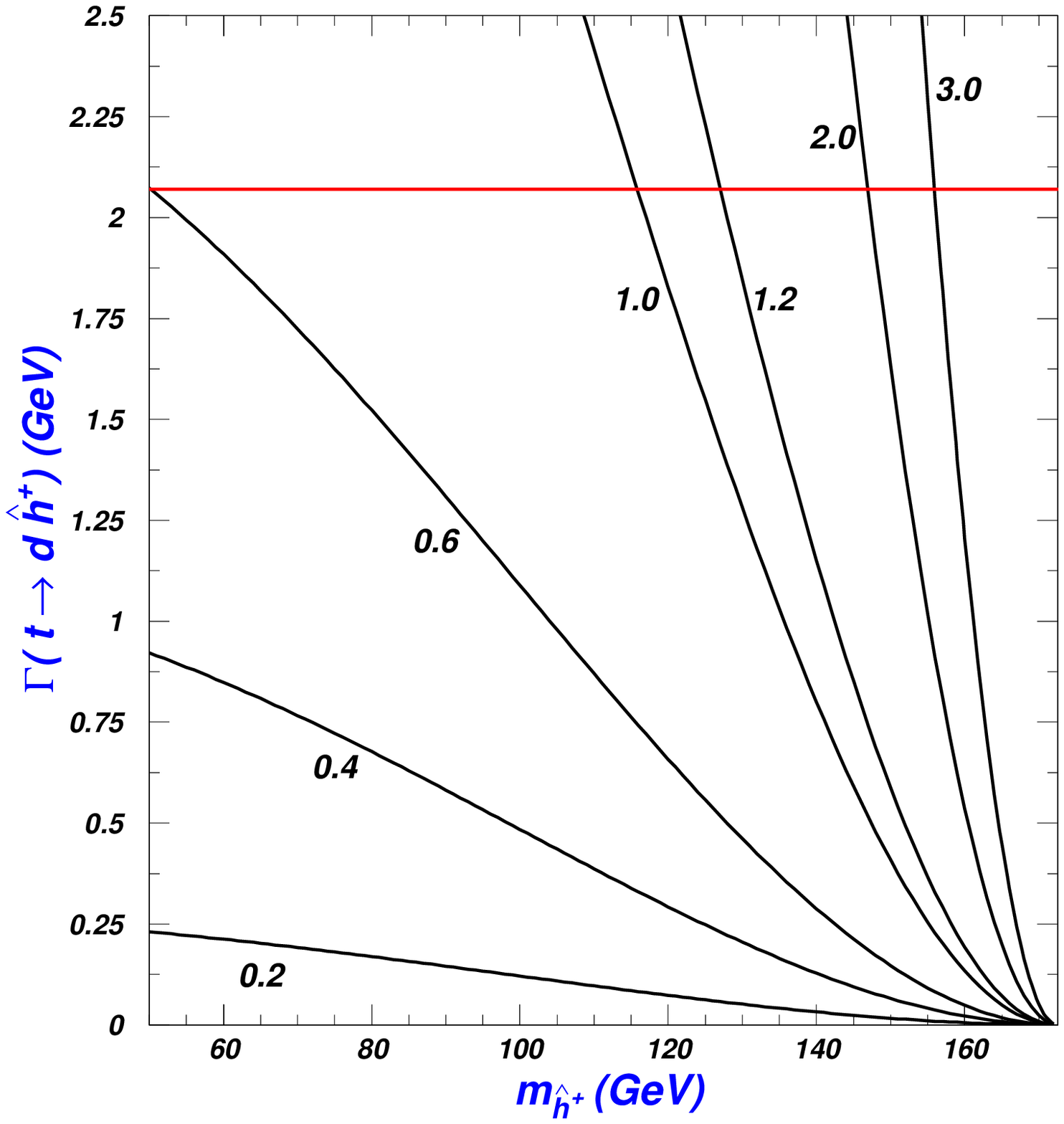,height=5.6cm}
\vspace{-1.5cm} \caption{For Case II, the new physics contributions
to $t\bar{t}$ production rates (normalized to SM values) and the
decay width of $t\to d\hat{h}^+$ versus $m_{\hat{h}^+}$. The numbers
on the curves denote the Yukawa coupling $y_2$. The horizontal lines
show the $2\sigma$ upper limits from the corresponding experimental
data.} \label{caseII}
\end{figure}

\begin{figure}[tb]
 \epsfig{file=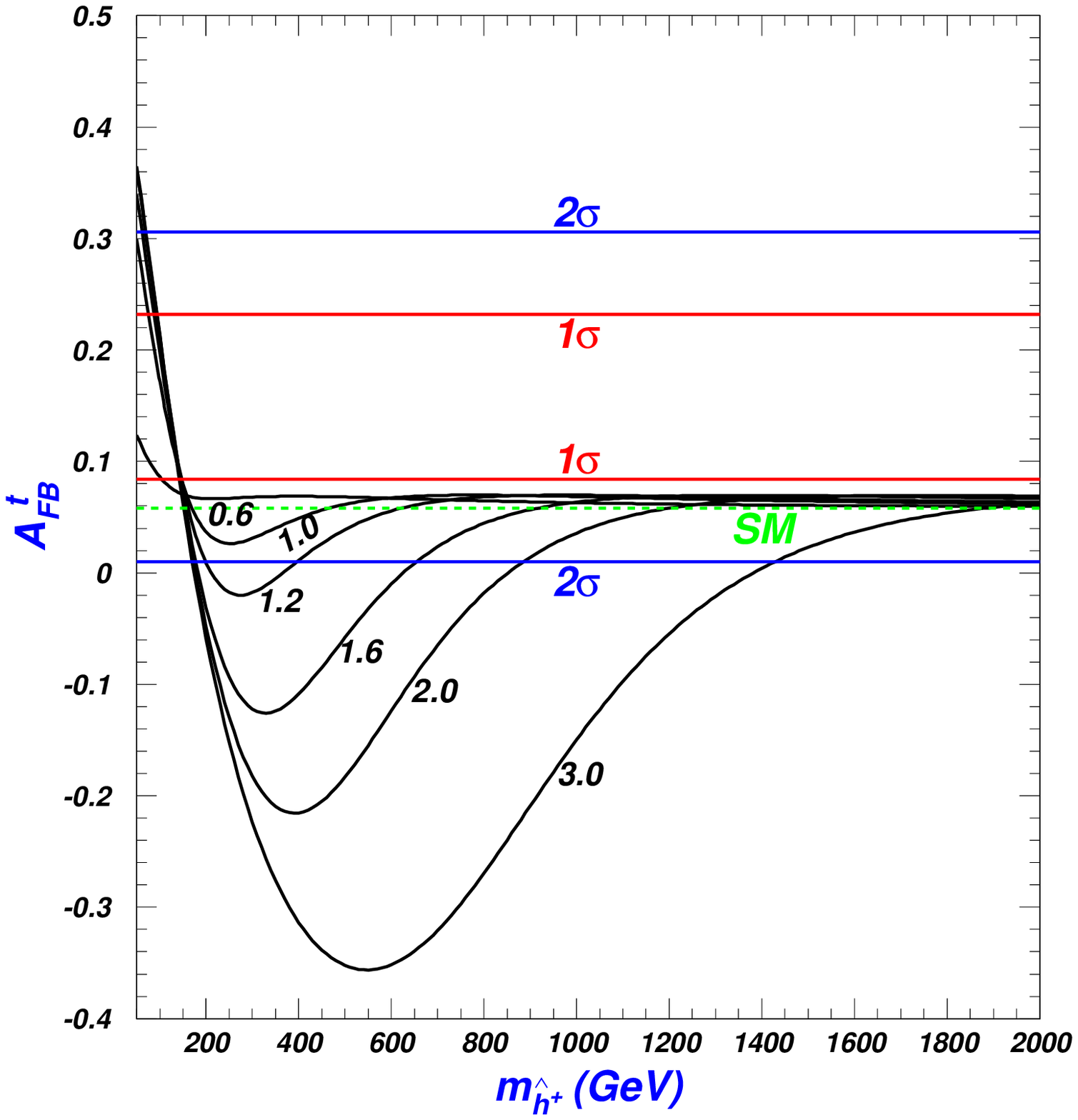,height=7.5cm}
 \epsfig{file=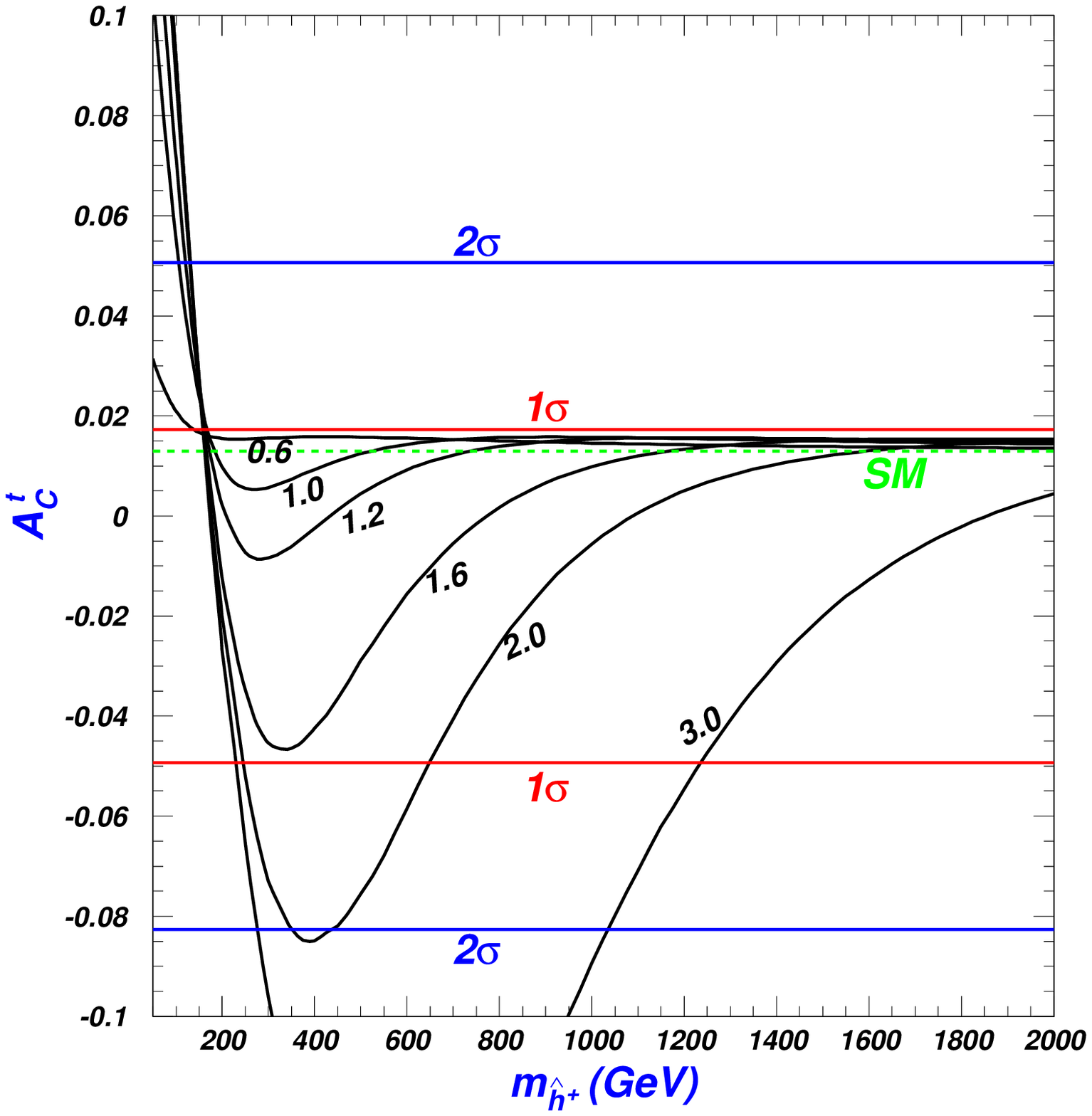,height=7.5cm}
\vspace{-.5cm}
\caption{Same as Fig.2, but for Case II.}
\label{IIafbc}
\end{figure}

\begin{figure}[tb]
 \epsfig{file=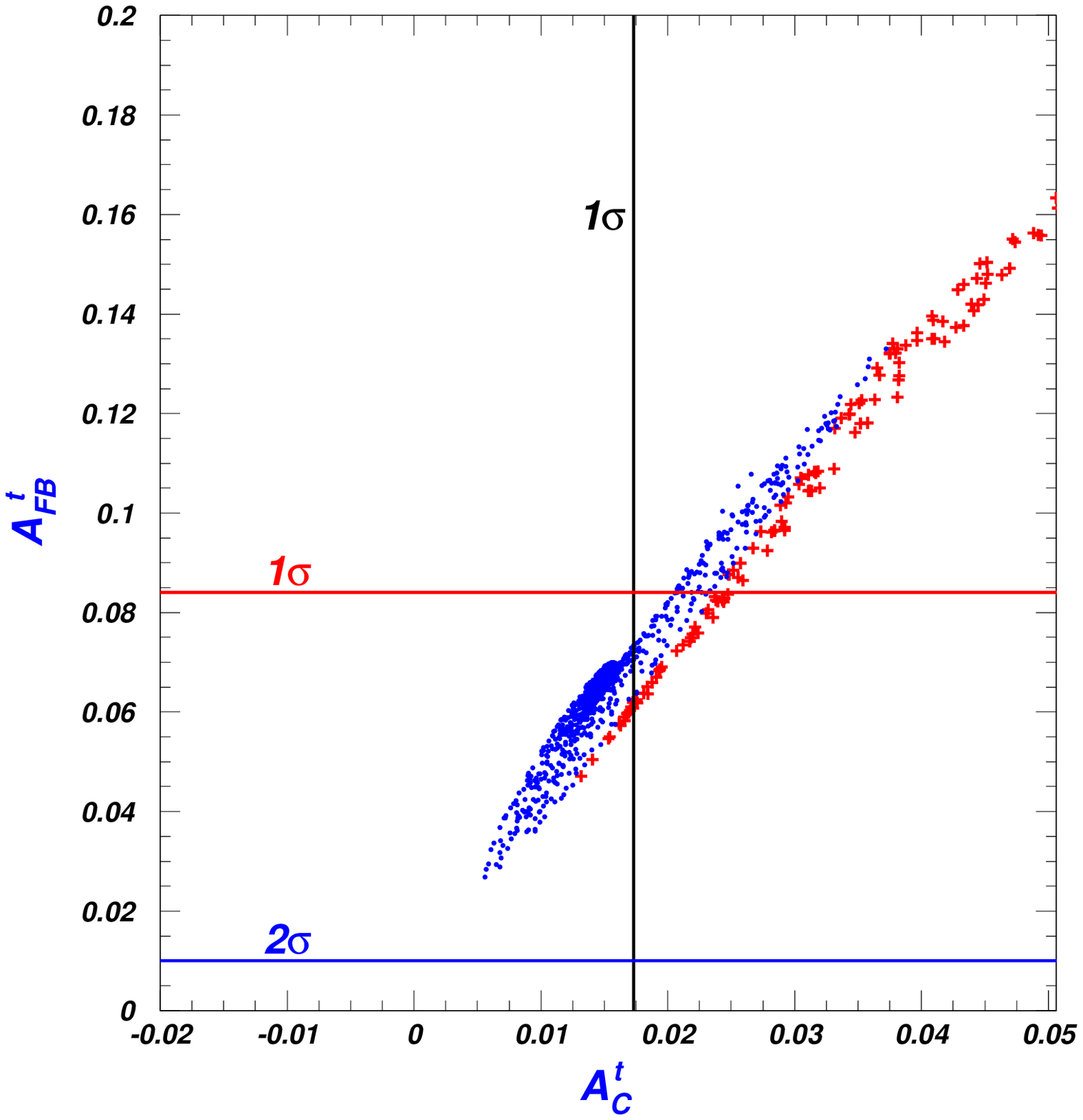,height=9.0cm}
\vspace{-.7cm}
\caption{Same as Fig.3, but for Case II}
\label{IIafblhc}
\end{figure}

In Fig. \ref{caseII}, we plot respectively the new physics
contributions to $t\bar{t}$ production at Tevatron and LHC
normalized to SM one, and the decay width of $t\to d\hat{h}^+$ for
Case II. Compared to Case I, the magnitude of
$\frac{\sigma^{NP}}{\sigma^{SM}}$ at Tevatron and LHC for Case II is
less sizable due to the smaller parton distribution function of $d$
quark. Therefore, the more broad region of the parameter space for
Case II is allowed by the related experimental data of top quark.
For example,  $m_{\hat{h}^+}$ is required to be larger than 180 GeV
(450 GeV) for $y_2 =1.0$ (1.6).

In Fig. \ref{IIafbc}, we plot the top quark forward-backward
asymmetry $A_{FB}^t$ at Tevatron and charge asymmetry $A_{C}^t$ at
LHC for Case II. The $A_{FB}^t$ can be enhanced sizably for the very
low values of $m_{\hat{h}^+}$, be negative in the intermediate
region and be outside the range of $1\sigma$ for the large ones
which differs from the Case I. The $A_{C}^t$ can be still within
$1\sigma$ in the most of parameter spaces.

In Fig. \ref{IIafblhc}, we scan the following parameter space,
$$0.1 \leq y_2 \leq 1.0,~~~100~ GeV \leq m_{\hat{h}^+} \leq 1000~GeV,$$ and
plot $A_{FB}^t$ versus $A_{C}^t$ under the constraints of the three
observables shown in Fig. \ref{caseII}. We find that the relative
large parameter space scanned is allowed by the three experimental
data of top quark. The $A_{FB}^t$ is outside the range of 1$\sigma$
for $A_{C}^t$ is within $1\sigma$, and reaches 0.13 for $A_{C}^t$
equals to 0.035 (at $1.5\sigma$). The measurement of $A_{FB}^t$ at
Tevatron is complementary to $A_{C}^t$ at LHC.

\subsection{Other discussions}

\begin{figure}[tb]
 \epsfig{file=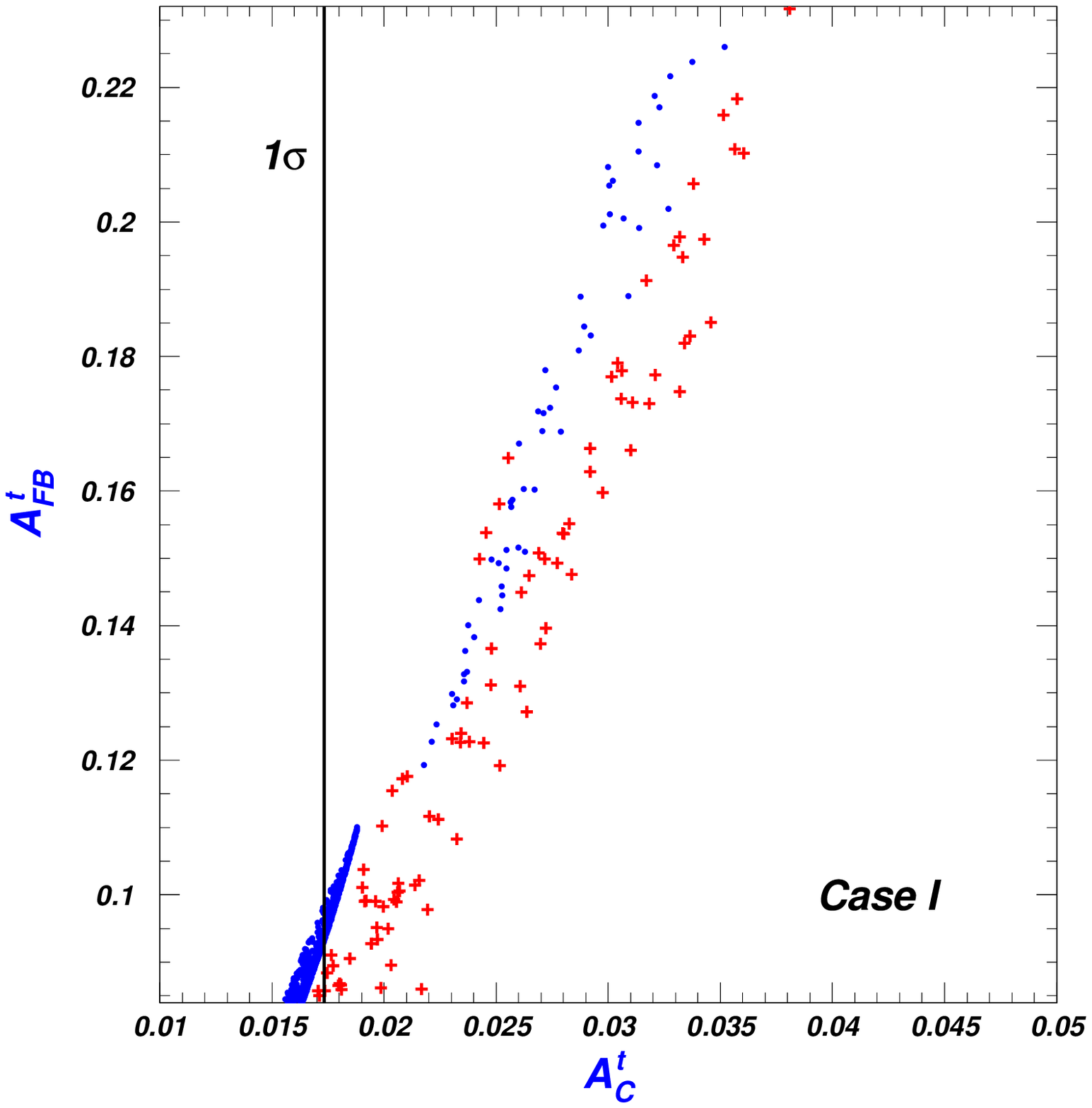,height=7.5cm}
 \epsfig{file=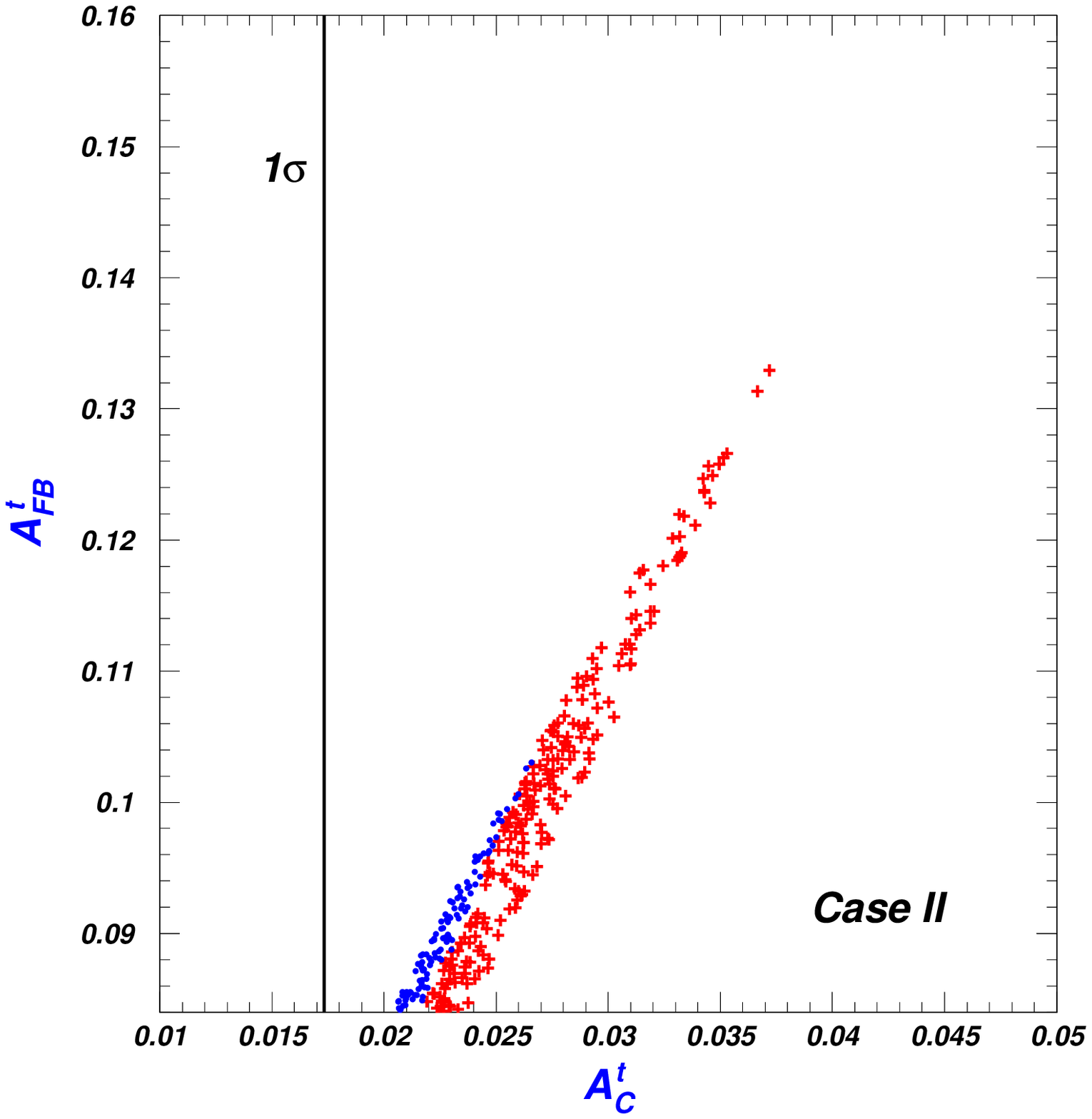,height=7.6cm}
\vspace{-.5cm} \caption{Top quark forward-backward asymmetry
$A_{FB}^t$ at Tevatron versus charge asymmetry
 $A_{C}^t$ at LHC. All the plots are in agreement with the constraints
 from the total width of top quark, $t\bar{t}$ production cross sections at
 Tevatron and LHC. The bullets (blue) and crosses (red) are
respectively allowed and excluded by the experimental data of the
$t\bar{t}$ invariant mass distribution at Tevatron.}
\label{invariant}
\end{figure}

The $t\bar{t}$ invariant mass distribution was measured by CDF, and
the results are presented in nine bins of $M_{t\bar{t}}$
\cite{09121447-22}, which does not give enough solid constraint on
this model since the QCD correction and cut efficiency may
significantly modify the shape of differential distribution
$d\sigma/d_{M_{t\bar{t}}}$ \cite{11040083,11040083-19,11040083-20}.
However, we will further examine the constraints of the invariant
mass distribution by requiring the differential cross section in
each bin to be within the $2\sigma$ regions of their experimental
values. We scan the $y_1$ ($y_2$) and $m_{\hat{h}^0}$
($m_{\hat{h}^+}$) in the region where the total width of top quark,
$t\bar{t}$ production cross sections at Tevatron and LHC are in
agreement with the corresponding constraints of experimental data.
We plot respectively $A_{FB}^t$ versus $A_{C}^t$ for Case I and Case
II in Fig. \ref{invariant}, where $A_{FB}^t$ is within $1\sigma$
range of the experimental value. From Fig. \ref{invariant}, we can
find that the constraints of $t\bar{t}$ invariant mass distribution
can further exclude some values of $A_{FB}^t$ and $A_C^t$. For Case
I, our previous conclusions are not changed. For Case II, some large
values of $A_{FB}^t$ are disfavored by the constraints of invariant
mass distribution.

\begin{figure}[tb]
 \epsfig{file=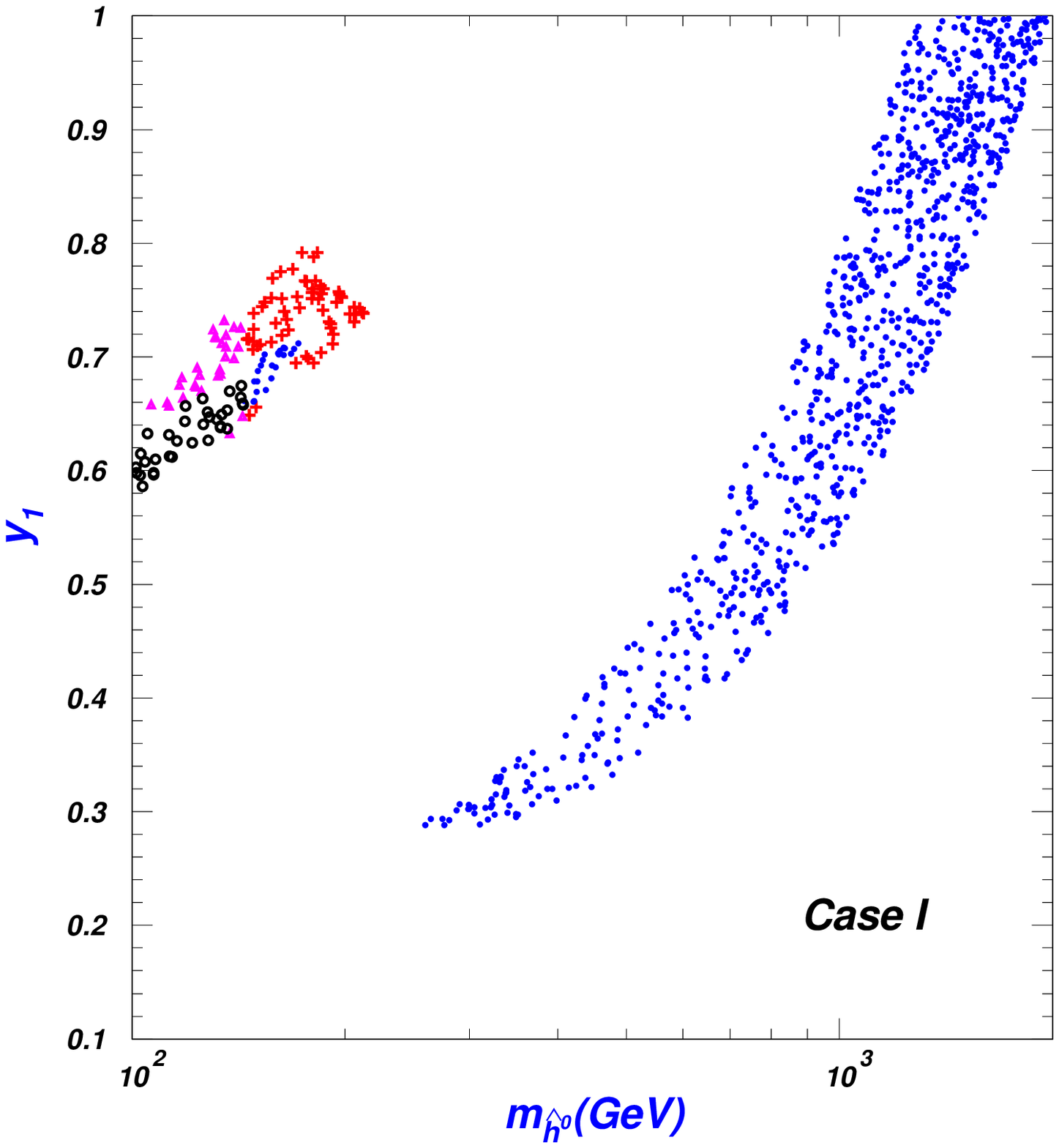,height=7.5cm}
 \epsfig{file=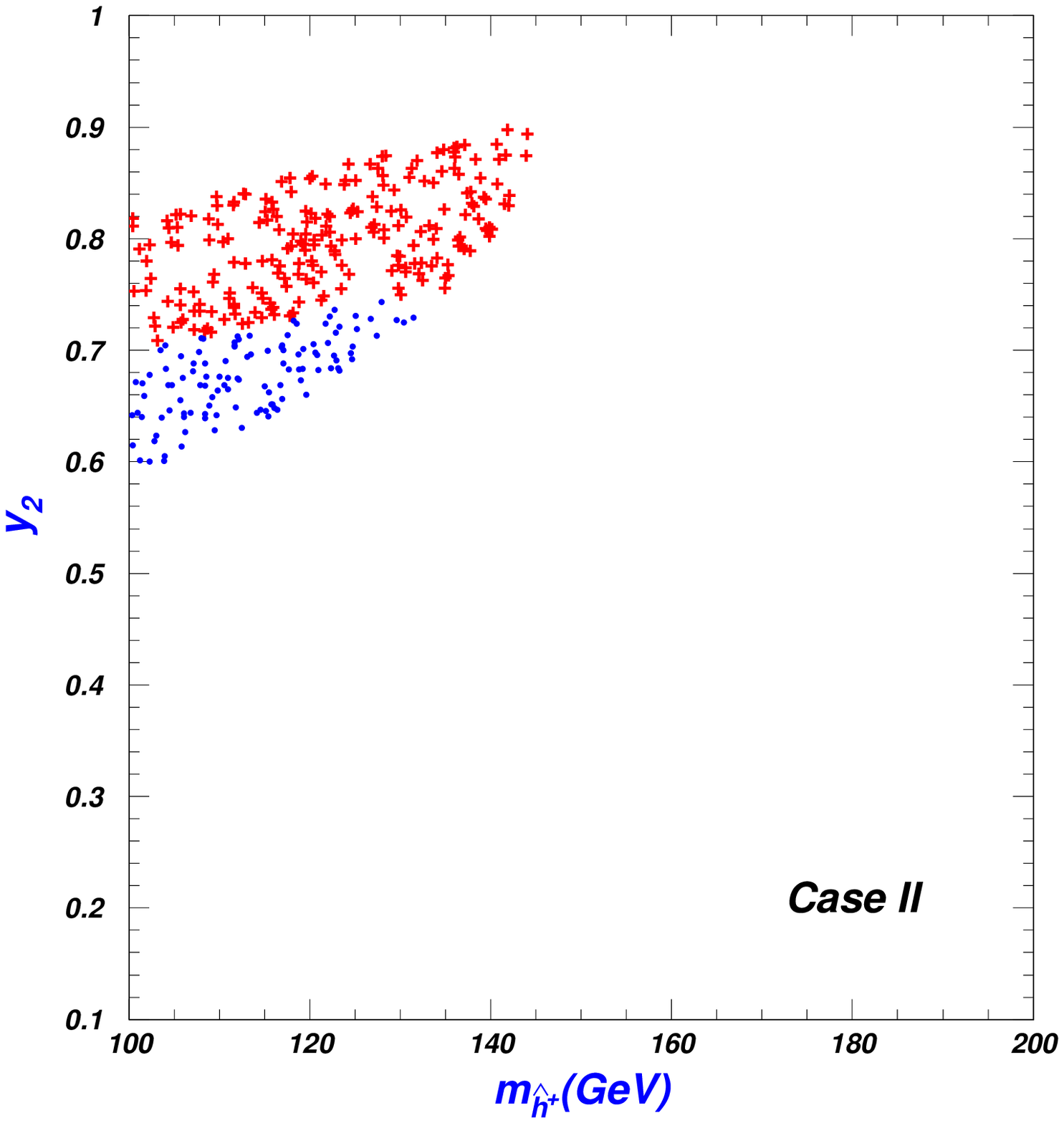,height=7.5cm}
\vspace{-.5cm} \caption{$y_1$ ($y_2$) and $m_{\hat{h}^0}$
($m_{\hat{h}^+}$) corresponding to Fig. \ref{invariant}. For the
bullets (blue) and crosses (red), $A_{FB}^t$ is within -1$\sigma$
range; For the circles (black) and triangle (pink), $A_{FB}^t$ is
within +1$\sigma$ range. The circles (black) and bullets (blue) are
allowed by the experimental data of the $t\bar{t}$ invariant mass
distribution at Tevatron; The crosses (red) and triangle (pink) are
excluded by this data.} \label{mhys}
\end{figure}

\begin{figure}[tb]
 \epsfig{file=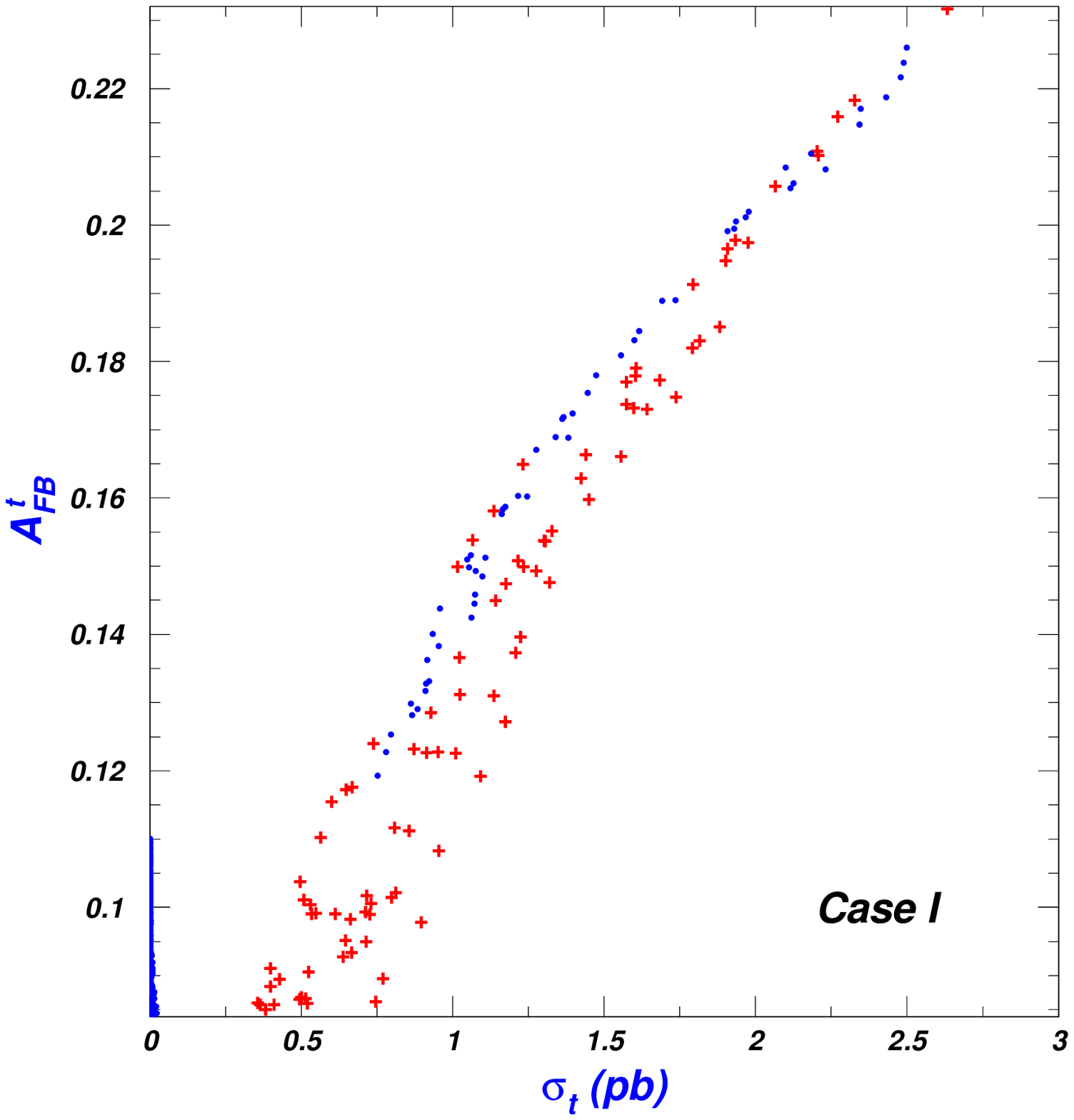,height=7.5cm}
 \epsfig{file=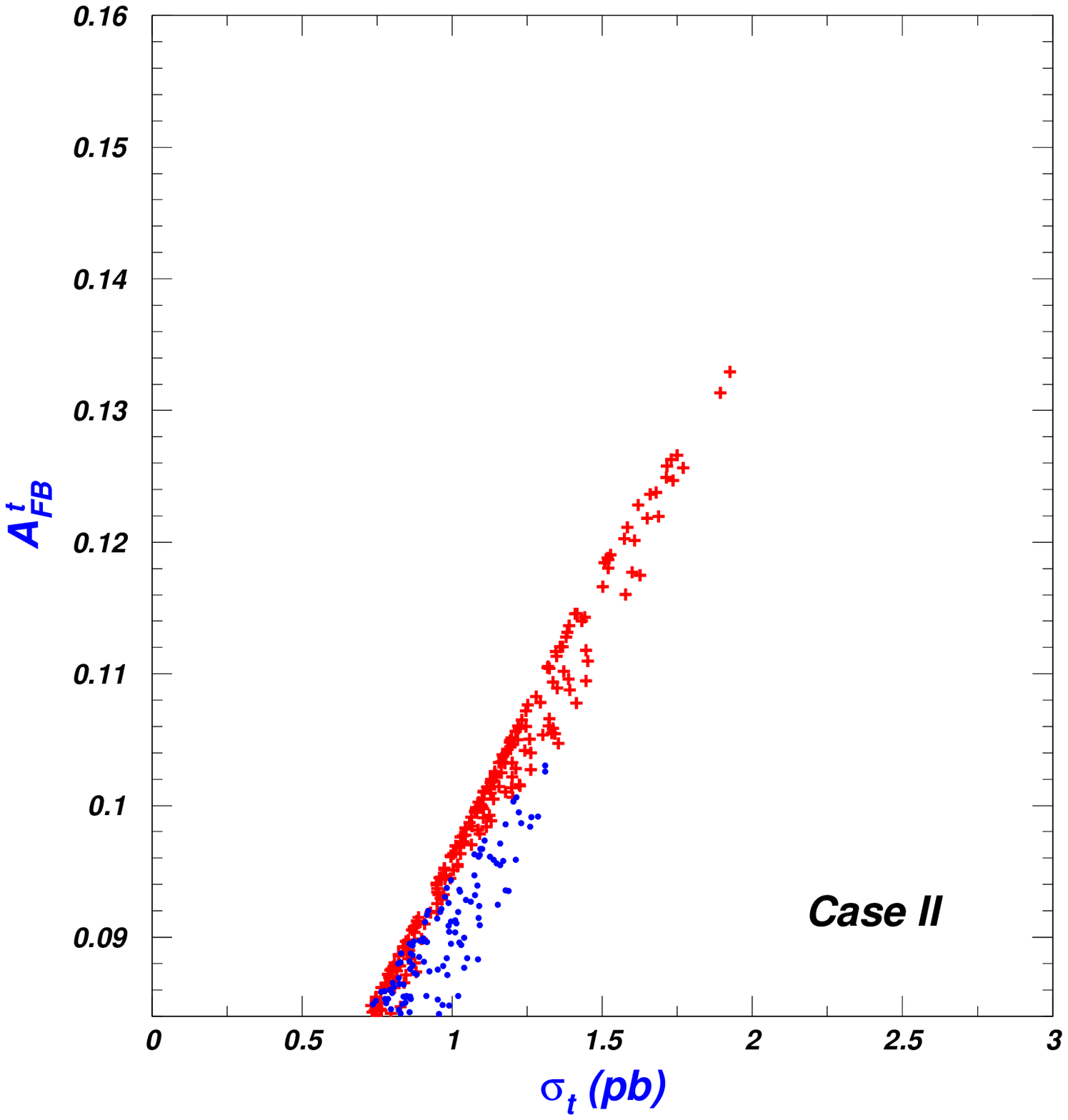,height=7.5cm}
\vspace{-.5cm} \caption{Same as Fig. \ref{invariant}, but for
$A_{FB}^t$ versus $\sigma_t$. $\sigma_t$ denotes the cross section
of the single top quark associated with the scalar production at
Tevatron. $\sigma_t\equiv\sigma\left(gu\to
t\hat{S}(\hat{A})\right)+\sigma\left(g\bar{u}\to
\bar{t}\hat{S}(\hat{A})\right)$ for Case I;
$\sigma_t\equiv\sigma\left(gd\to t\hat{h}^-
\right)+\sigma\left(g\bar{d}\to \bar{t}\hat{h}^+ \right)$ for Case
II.} \label{afbt}
\end{figure}

The values of $y_1$ ($y_2$) and $m_{\hat{h}^0}$ ($m_{\hat{h}^+}$)
corresponding to Fig. \ref{invariant} are shown in Fig. \ref{mhys}.
When 0.6 $\leq y_1 \leq 0.7$ (0.6 $\leq y_2 \leq 0.75$) and 100 GeV
$< m_{\hat{h}^0} < 200$ GeV (100 GeV$< m_{\hat{h}^+}<$ 140 GeV),
$A_{FB}^t$ is allowed to be within the $+1\sigma$ ($-1\sigma$) range
for Case I (Case II). In such parameter space, this model can fit
best the experimental data of $A_{FB}^t$. When $y_1 (y_2) =0.6$,
$\kappa_1 (\kappa_2)$ should be around 1.5 for $\Lambda=2\pi f$ and
3.0 for $\Lambda=4\pi f$ taking $\hat{f}=5f$ (see Eqs.
(\ref{yukhat1}) and (\ref{yukhat2})). Thus, an unnaturally large
$\kappa_1 (\kappa_2)$ is not necessary for $A_{FB}^t$ is within
$1\sigma$ range.

For Case I, $\hat{S}$ $(\hat{A})$ can decay into an up quark and an
up-type quark. For Case II, $\hat{h}^\pm$ can decay into a down
quark and an up-type quark. Except for the decay into top quark, the
other decays will be suppressed by the corresponding mixing matrix
element. For masses of these scalars are much larger than top quark
mass, their total widths can reach the half of the masses taking
$y_1=y_2=1$, respectively. We find that the value of $A_{FB}^t$ is
not changed sizably when varying the width from zero GeV to the half
of scalar mass, especially for that $A_{FB}^t$ is within $+1\sigma$
range for Case I and within $1\sigma$ range for Case II. The reason
is that the widths of these scalars are very small for such values
of $A_{FB}^t$, which can be derived according to the parameters
shown in Fig. \ref{mhys}.

The D0 has recently measured single top quark production cross
section at Tevatron by requiring one $b$-jet in the final states and
obtained $\sigma(p\bar{p}\to tqb+X)=2.90\pm0.59$ pb \cite{11052788},
where $q$ is a light quark. The experimental value is in agreement
with the SM t-channel $tbq$ result of $2.26\pm0.12$ pb. For Case I
and Case II, the single top can be produced by the process $gu\to
t\hat{S}~(\hat{A})$ and $gd\to t \hat{h}^-$, respectively. In Fig.
\ref{afbt}, we plot the $A_{FB}^t$ versus the cross sections of the
single top quark associated with the scalar production at Tevatron
for Case I and Case II. We find that the cross sections can be over
1 pb when $A_{FB}^t$ is larger than 0.15 for Case I and 0.1 for Case
II, respectively. However, given that $\hat{S}$, $\hat{A}$ and
$\hat{h}^\pm$ can not decay into a bottom quark, this constraint is
not suitable for our model due to the lack of $b$-jet in the final
states. A dedicated study is required in order to establish the
applicability of the single top measurements at the Tevatron to our
model.

In the LRTH model, there exist additional heavy gauge bosons from
the $SU(2)_R$ symmetric sector, dubbed $W_H^{\pm}$ and $Z_H$, which
can also contribute to the top quark forward-backward asymmetry. In
this model, the $SU(2)_{L,R}$ coupling constants $g_L$ and $g_R$ are
identical. The experimental limits favor that the quark mixing
matrices in the left- and right-handed sectors are the same
\cite{phlrth}. For this case, Ref. \cite{11080998} shows that the
value of $A_{FB}^t$ produced by $W_H^{\pm}$ and $Z_H$ is much
smaller than the experimental value. Compared with the contributions
of $\hat{h}^{\pm}$, $\hat{S}$ and $\hat{A}$, their contributions can
be ignored safely.

\section{Conclusion}
In the framework of left-right twin Higgs model we introduced a new
Yukawa interaction for the doublet $\hat{h}$, which leads that the
lightest neutral particle of $\hat{h}$ can no longer be the dark
matter candidate. Such new Yukawa interaction was found to sizably
contribute to the top quark forward-backward asymmetry $A_{FB}^t$ at
the Tevatron. Under the constraints from the related experimental
data of top quark, we found that the Tevatron $A_{FB}^t$ can be
explained while the LHC charge asymmetry $A_{C}^t$ measurement can
also be satisfied.

Although explaining $A_{FB}^t$ by extending Higgs sector has been
studied in some papers, most of them do not propose a realistic
model. By introducing the new Yukawa interaction, we make the LRTH
to be a realistic one, which can solve the hierarchy problem in
addition to $A_{FB}^t$. Besides, the degeneracy masses of $\hat{S}$
and $\hat{A}$ can naturally avoid the strong constraints of the
same-sign top pair production at LHC, leading $A_{FB}^t$ to reach
0.24.

\section*{Acknowledgment}
We thank Manuel Perez-Victoria for helpful comment. This work was
supported in part by the National Natural Science Foundation of
China (NNSFC) under grant Nos. 11105116 and 11005089, 10725526,
10821504 and 10635030, and by the Project of Knowledge Innovation
Program (PKIP) of Chinese Academy of Sciences under grant No.
KJCX2.YW.W10.


\begin{thebibliography}{99}

\bibitem{11100014-1} CDF Collaboration, \PRD83, 112003 (2011).

\bibitem{11100014-2} D0 Collaboration, \PRD84, 112005 (2011).

\bibitem{11084005-4¨C7} J. H. Kuhn and G. Rodrigo,
\PRL81, 49 (1998); J. H. Kuhn and G. Rodrigo, \PRD59, 054017 (1999);
M. T. Bowen, S. D. Ellis and D. Rainwater, \PRD 73, 014008 (2006);
O. Antunano, J. H. Kuhn and G. Rodrigo, \PRD77, 014003 (2008).

\bibitem{11066051} V. Ahrens, A. Ferroglia, M. Neubert, B. D. Pecjak,
L. L. Yang, \PRD84, 074004 (2011).


\bibitem{afb-s-channel}   
  P.~Ferrario, G.~Rodrigo, \PRD80, 051701 (2009);
  P.~Ferrario and G.~Rodrigo, JHEP {\bf 1002}, 051 (2010);
  P.~H.~Frampton {\it et al.}, \PRD683, 294 (2010);
  M.~V.~Martynov, A.~D.~Smirnov, Mod.\ Phys.\ Lett.\  A {\bf 25}, 2637 (2010);
  R. S. Chivukula {\it et al.}, \PRD82, 094009 (2010);
  Y.~Bai {\it et al.}, JHEP {\bf 1103}, 003 (2011);
  A.~Djouadi {\it et al.}, \PRD82, 071702 (2010);
  K.~Kumar  {\it et al.}, JHEP {\bf 1008}, 052 (2010);
  G.~Burdman {\it et al.},  Phys.\ Rev.\  D {\bf 83}, 035012 (2011);
  E.~Alvarez {\it et al.}, JHEP {\bf 1105}, 070 (2011);
  C.~Delaunay {\it et al.}, arXiv:1101.2902;
  M.~Bauer {\it et al.}, JHEP 1011, 039 (2010);
  C.~H.~Chen {\it et al.}, Phys.\ Lett.\  B {\bf 694}, 393 (2011);
  R.~Foot, Phys.\ Rev.\  D {\bf 83}, 114013 (2011);
  A.~Djouadi {\it et al.}, Phys.\ Lett.\  B {\bf 701}, 458 (2011);
  R.~Barcelo {\it et al.}, arXiv:1105.3333;
  G.~M.~Tavares and M.~Schmaltz, arXiv:1107.0978;
  E.~Alvarez {\it et al.}, arXiv:1107.1473;
  E.~Gabrielli, M.~Raidal, arXiv:1106.4553;
  H.~Wang {\it et al.}, arXiv:1107.5769;
  G.~Z.~Krnjaic, arXiv:1109.0648;
  H.~Davoudiasl, T.~McElmurry, A.~Soni, arXiv:1108.1173;
 E.~L.~Berger {\it et al.}, arXiv:1111.3641;
X.-P. Wang et al., \PRD83, 115010 (2011);
J. A. Aguilar-Saavedra, M. Perez-Victoria, \PLB705, 228-234 (2011).


\bibitem{afb-tu-channel}  
  K.~Cheung {\it et al.}, \PLB682, 287 (2009);
  S.~Jung,  {\it et al.},
  Phys.\ Rev.\  D {\bf 81}, 015004 (2010);
  V. Barger {\it et al.}, \PRD81, 113009 (2010);
  I.~Dorsner {\it et al.}, \PRD81, 055009 (2010);
  A.~Arhrib, R.~Benbrik, C.~H.~Chen, \PRD82, 034034 (2010);
  G.~Rodrigo, P.~Ferrario, Nuovo Cim. C {\bf 33}, 04 (2010);
  J. Cao {\it et al.}, \PRD81, 014016 (2010);  \PRD83, 034024 (2011);
  \PRD84, 074001 (2011); arXiv:1109.6543;
  S.~Jung, A.~Pierce, J.~D.~Wells,
  Phys.\ Rev.\  D {\bf 83}, 114039 (2011);
  B.~Bhattacherjee {\it et al.}, \PRD83, 091501 (2011);
  K.~M.~Patel, P.~Sharma, \JHEP1104, 085 (2011);
  M.~R.~Buckley,  {\it et al.}, \PRD83, 115013 (2011);
  G.~Isidori and J.~F.~Kamenik, \PLB700, 145 (2011);
  E.~R.~Barreto {\it et al.}, \PRD83, 054006 (2011);
  A.~Rajaraman, Z.~E.~Surujon, T.~M.~P.~Tait, arXiv:1104.0947;
  M.~I.~Gresham {\it et al.}, arXiv:1107.4364;
  Y.~Cui {\it et al.}, arXiv:1106.3086;
  M.~Duraisamy, A.~Rashed, A.~Datta, arXiv:1106.5982;
  B.~Grinstein,  {\it et al.}, arXiv:1108.4027;
  D.~Kahawala, D.~Krohn, M.~J.~Strassler, arXiv:1108.3301;
  P.~Ko, Y.~Omura, C.~Yu, arXiv:1108.4005;
  S.~K.~Gupta, arXiv:1011.4960;
  E.~L.~Berger {\it et al.}, Phys.\ Rev.\ Lett.\  {\bf 106}, 201801 (2011); arXiv:1109.3202;
  K.~Cheung and T.~C.~Yuan, Phys.\ Rev.\  D {\bf 83}, 074006 (2011);
  Z.~Ligeti, G.~M.~Tavares, M.~Schmaltz, JHEP {\bf 1106}, 109 (2011).
  M.~I.~Gresham, I.~W.~Kim, K.~M.~Zurek,
  Phys.\ Rev.\  D {\bf 83}, 114027 (2011);
  J.~F.~Kamenik, J.~Shu, J.~Zupan, arXiv:1107.5257;
  S.~Westhoff, arXiv:1108.3341;
  K. Yan {\it et al.}, arXiv:1110.6684.

\bibitem{11040083}  J.~Shu, K.~Wang, G.~Zhu, \PRD85, 034008 (2012).


\bibitem{11040083-19} B. Xiao, Y.-k. Wang, S. -h. Zhu, \PRD 82,
034026 (2010).

\bibitem{11080998} M. Frank, A. Hayreter, I. Turan,
\PRD84, 114007 (2011).

\bibitem{afbdefi} Q.-H.~Cao {\it et al.}, \PRD 81, 114004
(2010).

\bibitem{11074350} K. Blum, Y. Hochberg, Y. Nir, \JHEP1110, 124 (2011).

\bibitem{09113237} J. Shu, T. M. P. Tait, K. Wang, \PRD81, 034012
(2010).

\bibitem{afb-eft}  
   D.~W.~Jung {\it et al.}, \PLB691, 238 (2010);
  arXiv:1012.0102;
  C. Zhang, S. Willenbrock, arXiv:1008.3869;
  J.~A.~Aguilar-Saavedra, \NPB843, 638 (2011);
  Nucl.\ Phys.\  B {\bf 812}, 181 (2009);
  C.~Degrande {\it et al.}, arXiv:1010.6304;
  K.~Blum {\it et al.}, arXiv:1102.3133;
  C.~Delaunay {\it et al.}, arXiv:1103.2297;
  C.~Degrande {\it et al.}, arXiv:1104.1798;
  D.~Y.~Shao {\it et al.}, arXiv:1107.4012;
  J.~A.~Aguilar-Saavedra, M.~Perez-Victoria, Phys.\ Lett.\  B {\bf 701}, 93 (2011);
   \JHEP1105, 034 (2011).

\bibitem{afb-eft-2}
  J. A. Aguilar-Saavedra, M. Perez-Victoria, \JHEP1109, 097 (2011).

\bibitem{twinhiggs}
Z. Chacko, H. S. Goh, and R. Harnik, \PRL96, 231802 (2006); R.
Barbieri, T. Gregoire, and L. J. Hall, hep-ph/0509242; Z. Chacko, Y.
Nomura, M. Papucci, G. Perez, \JHEP01, 126 (2006); R. Foot, R. R.
Volkas, \PLB645, 75 (2007); A. Falkowski, S. Pokorski, M. Schmaltz,
\PRD74, 035003 (2006); S. Chang, L. J. Hall, N. Weiner, \PRD75,
035009 (2007).

\bibitem{lrth} Z. Chacko, H. S. Goh, R. Harnik, \JHEP0601, 108 (2006).

\bibitem{phlrth} H. S. Goh, S. Su, \PRD75, 075010 (2007).

\bibitem{lrthdm} E. M. Dolle, S. Su, \PRD77, 075013 (2008);
                 L. Wang, J. M. Yang, \JHEP1005, 024 (2010).

\bibitem{ac-np}
  J.~L.~Hewett {\it et al.}, \PRD84, 054005 (2011);
  J.~F.~Arguin, M.~Freytsis and Z.~Ligeti, \PRD84, 071504 (2011).

\bibitem{lacafbdefi} http://cdsweb.cern.ch/record/1369205/files/TOP-11-014-pas.

\bibitem{atlas-ac}
http://cdsweb.cern.ch/record/1372916/files/ATLAS-CONF-2011-106.

\bibitem{11070841-46} T. Aaltonen et al. [CDF Collaboration], CDF note 9913.

\bibitem{11070841-47} U. Langenfeld, S. Moch, P. Uwer, \PRD80, 054009 (2009).

\bibitem{11070841-48} V. Ahrens,  et al., \JHEP1009, 097 (2010).

\bibitem{11101027} ATLAS Collaboration, arXiv:1110.1027.

\bibitem{11055661} CMS Collaboration, \JHEP1107, 049 (2011).

\bibitem{top11008pas-9} V. Ahrens et al., \JHEP09, 097 (2010).

\bibitem{11074350-49} V. M. Abazov et al. [D0 Collaboration ], \PRL106, 022001 (2011).

\bibitem{0701071} D.-W. Jung, J. Y. Lee, hep-ph/0701071.

\bibitem{11062142} CMS Collaboration, \JHEP1108, 005 (2011).

\bibitem{cteq} J. Pumplin et al., \JHEP0602, 032 (2006).

\bibitem{09121447-22} CDF Collaboration], \PRL102, 222003 (2009).


\bibitem{11040083-20} M. I. Gresham, I. -W. Kim, K. M. Zurek,
\PRD83, 114027 (2011).

\bibitem{11052788} D0 Collaboration, \PLB705, 313 (2011).


\end{thebibliography}
\end{document}